\documentclass[submitting]{nst}

\usepackage{subfigure,dcolumn}
\usepackage[T2A,T1]{fontenc}
\usepackage[russian,english]{babel}
\usepackage{color,xcolor} 
\usepackage{array, makecell}
\usepackage{lineno,hyperref}

\usepackage{listings}
\begin{document}

\title{Event vertex and time reconstruction in large-volume liquid scintillator detectors}\thanks{This work was supported by the National Natural Science Foundation of China (Nos. 11805294 and 11975021), the China Postdoctoral Science Foundation (2018M631013), the Strategic Priority Research Program of Chinese Academy of Sciences (XDA10010900), the Fundamental Research Funds for the Central Universities, Sun Yat-sen University (19lgpy268), and in part by the CAS Center for Excellence in Particle Physics (CCEPP).}

\author{Zi-Yuan Li}
\affiliation{School of Physics, Sun Yat-Sen University, Guangzhou 510275, China}
\author{Yu-Mei Zhang}
\email[Yu-Mei Zhang, ]{zhangym26@mail.sysu.edu.cn}
\affiliation{Sino-French Institute of Nuclear Engineering and Technology, Sun Yat-Sen University, Zhuhai 519082, China}
\author{Zhen Qian}
\affiliation{School of Physics, Sun Yat-Sen University, Guangzhou 510275, China}
\author{Shu Zhang}
\affiliation{School of Physics, Sun Yat-Sen University, Guangzhou 510275, China}
\author{Kai-Xuan Huang}
\affiliation{School of Physics, Sun Yat-Sen University, Guangzhou 510275, China}
\author{Guo-Fu Cao}
\affiliation{Institute of High Energy Physics, Chinese Academy of Sciences, Beijing 100049, China}
\author{Zi-Yan Deng}
\affiliation{Institute of High Energy Physics, Chinese Academy of Sciences, Beijing 100049, China}
\author{Gui-Hong Huang}
\affiliation{Institute of High Energy Physics, Chinese Academy of Sciences, Beijing 100049, China}
\author{Wei-Dong Li}
\affiliation{Institute of High Energy Physics, Chinese Academy of Sciences, Beijing 100049, China}
\author{Tao Lin}
\affiliation{Institute of High Energy Physics, Chinese Academy of Sciences, Beijing 100049, China}
\author{Liang-Jian Wen}
\affiliation{Institute of High Energy Physics, Chinese Academy of Sciences, Beijing 100049, China}
\author{Miao Yu}
\affiliation{School of Physics and Technology, Wuhan University, Wuhan 430072, China}
\author{Jia-Heng Zou}
\affiliation{Institute of High Energy Physics, Chinese Academy of Sciences, Beijing 100049, China}
\author{Wu-Ming Luo}
\email[Wu-Ming Luo, ]{luowm@ihep.ac.cn}
\affiliation{Institute of High Energy Physics, Chinese Academy of Sciences, Beijing 100049, China}
\author{Zheng-Yun You}
\email[Zheng-Yun You, ]{youzhy5@mail.sysu.edu.cn}
\affiliation{School of Physics, Sun Yat-Sen University, Guangzhou 510275, China}

\begin{abstract}
Large-volume liquid scintillator detectors with ultra-low background levels have been widely used to study neutrino physics and search for dark matter. Event vertex and event time are not only useful for event selection but also essential for the reconstruction of event energy.	
	In this study, four event vertex and event time reconstruction algorithms using charge and time information collected by photomultiplier tubes were analyzed comprehensively. The effects of photomultiplier tube properties were also investigated. The results indicate that the transit time spread is the main effect degrading the vertex reconstruction, while the effect of dark noise is limited. In addition, when the event is close to the detector boundary, the charge information provides better performance for vertex reconstruction than the time information.

\end{abstract}


\maketitle

\section{Introduction}
	Liquid scintillators (LSs) have widely been used as detection medium for neutrinos in experiments such as Kamioka Liquid Scintillator Antineutrino Detector (KamLAND)~\cite{kamland}, Borexino~\cite{borexino}, Double Chooz~\cite{doublechooz}, Daya Bay~\cite{dyb}, and Reactor Experiment for Neutrino Oscillation (RENO)~\cite{reno}. KamLAND revealed a large mixing angle (LMA) solution for solar neutrino oscillations. Borexino confirmed the Mikheyev-–Smirnov-–Wolfenstein (MSW) LMA~\cite{{lmamsw}} model in the sub-MeV region for solar neutrino oscillations. Double Chooz, Daya Bay, and RENO reported nonzero measurements for the mixing angle $\theta_{13}$. The size of such detectors varies from hundreds to thousands of cubic meters. Large-volume liquid scintillator detectors are widely used in the next generation of neutrino experiments, aiming to solve problems such as neutrinoless double-beta decay (SNO+ at the Sudbury Neutrino Observatory~\cite{sno+}) and neutrino mass ordering (Jiangmen Underground Neutrino Observatory (JUNO)~\cite{junophysics}).
	
	The sensitivity of these experiments is limited by the energy resolution, detector volume, and detector background. These detectors typically contain a fiducial volume, where the signal-to-noise ratio is maximal. To distinguish between events occurring in the fiducial and non-fiducial regions, the event vertex is reconstructed using the charge and time distribution of photons collected by the photomultiplier tubes (PMTs). Most importantly, to achieve a high energy resolution, an accurate vertex is essential to correct for energy non-uniformity. In addition, the event vertex and event time information can also be used for particle identification, direction reconstruction, event classification, and other purposes. 
	
	A previous study~\cite{Liu_2018} investigated vertex reconstruction with time information in JUNO, without discussing the event time reconstruction, dark noise effect, and the improvement based on the charge information when the event is close to the detector boundary. The aim of this study was to analyze vertex and time reconstruction for point-like events in JUNO under more realistic conditions. The main contributions of this study are as follows:
	\begin{enumerate}
		\item[$*$] The event time was reconstructed to provide the start time information of the event, which was important for event alignment, event correlation, etc. 
		\item[$*$] The dark noise from PMTs was considered and its effect on the vertex reconstruction was properly controlled.		
		\item[$*$] Two types of large PMTs were considered and handled separately mainly because of the difference in transit time spread (TTS).
		\item[$*$] An algorithm to provide a more accurate initial vertex value was developed to improve the performance, especially at the detector boundary region.
		\item[$*$] An algorithm employing charge information to reconstruct the event vertex at the detector boundary was developed.
	\end{enumerate}	
	
	The remainder of this paper is organized as follows. In Sec.~\ref{sec:junodetector}, a brief introduction of the JUNO detector and the configurations of the PMTs used in this study is provided. In Sec.~\ref{sec:op}, the optical processes are described and a simple optical model is introduced. In Sec.~\ref{sec:initvalue}, two simple algorithms that can quickly provide initial values are compared. In Secs.~\ref{sec:tlh} and~\ref{sec:clh}, two complex algorithms that provide relatively good performance are introduced in detail. Finally, a performance summary, the discussion, and the conclusions are provided in Secs.~\ref{sec:summary},~\ref{sec:discussion}, and~\ref{sec:conclusion}, respectively.
	
\section{The JUNO detector}
\label{sec:junodetector}

	\begin{figure}[!ht]
		\centering
		\includegraphics[width=0.5\textwidth]{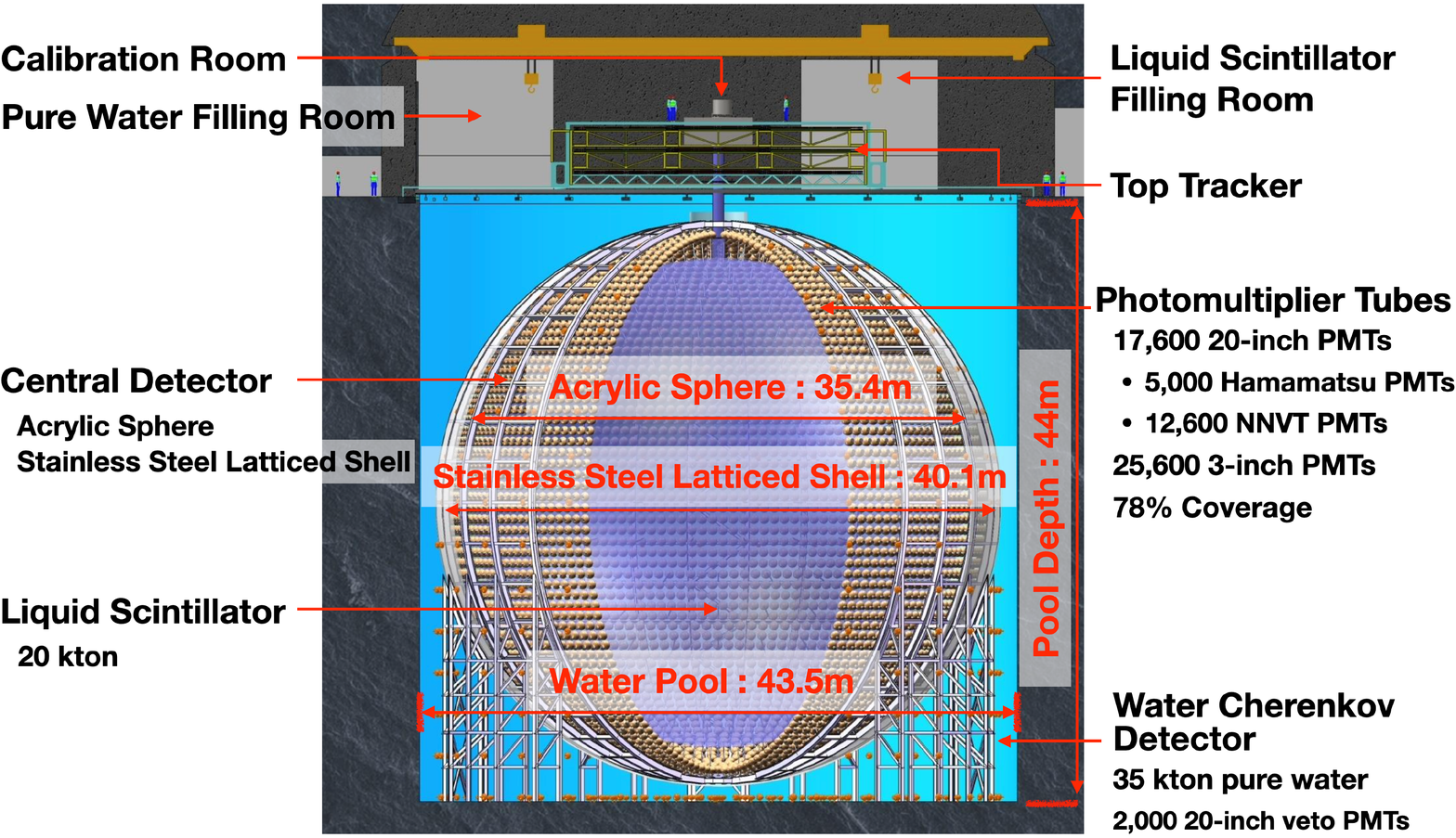}{\centering}									
		\caption{Schematic of the JUNO detector.}
		\label{fig:junodetector}			
	\end{figure}   

	A schematic of the JUNO detector is shown in Fig.~\ref{fig:junodetector}. The central detector (CD) is the main part of the JUNO detector assembly, with an acrylic sphere with a diameter of 35.4~m as inner layer. The acrylic sphere is supported by a stainless steel latticed shell with a diameter of 40.1~m and filled with approximately 20,000 tons of LS as target for neutrino detection. The composition of the LS includes linear alkylbenzene (LAB) as solvent, 2,5-diphenyloxazole (PPO) as fluor, and p-bis-(o-methylstyryl)-benzene (bis-MSB) as wavelength shifter. To collect photons, the CD is surrounded by approximately 17,600 20-inch PMTs, including 5000 Hamamatsu R12860 PMTs, 12,600 North Night Vision Technology (NNVT) GDG-6201 PMTs, and approximately 25,600 3-inch XP72B22 PMTs. Around the CD, a 2.5-m-thick water pool is used to shield external radioactivity from the surrounding rocks and is combined with approximately 2,000 20-inch GDG-6201 PMTs to serve as a water Cherenkov detector to veto cosmic ray muons. A top tracker detector, consisting of plastic scintillators, located above the water pool is used for the identification and veto of muon tracks. A more detailed description of the JUNO detector can be found in Refs.~\cite{junocdr,junophysics}.
	
	The main factors affecting the reconstruction of the event vertex and time include the TTS and dark noise of the PMTs. In this study, only the 20-inch PMTs of the CD were used for reconstruction. The number and parameters of the Hamamatsu and NNVT PMTs are summarized in Table~\ref{tab:pmttype}~\cite{pmtsys}. In principle, the 3-inch PMTs with $\sigma \simeq 1.6$~ns TTS could also be included to improve the reconstruction performance; however, these PMTs~\cite{3inchpmt} were not considered in this study because of their small photon detection coverage ($\sim$3\%)~\cite{gdml,gdml2}.
	
	\begin{table}[!htb]
		\centering
    		\caption{Number and parameters of the PMTs used in the reconstruction. The TTS and dark noise rate are the mean values of the distribution measured during the mass testing. However, these are not the final values for JUNO.}		
		\begin{tabular*}{8cm} {@{\extracolsep{\fill} } lccc}
 	       		\toprule
        		 	Company 		& Number		& TTS ($\sigma$)		& Dark noise rate \\
			\midrule
        			Hamamatsu	& 5,000 		& 1.15~ns 	& 15~kHz \\
        			NNVT		& 12,600 		& 7.65~ns 	& 32~kHz \\
        			\bottomrule
		\end{tabular*}
    		\label{tab:pmttype}
	\end{table}

\section{Optical processes}
\label{sec:op}
	When a charged particle deposits energy in the scintillator, the solvent enters an excited state and transfers energy to the fluor in a non-radiative manner. Scintillation photons are then emitted along the particle track through the radiative de-excitation of the excited fluor within a limited time. The emitted scintillation photons can undergo several different processes while propagating through a large LS detector. At short wavelengths ($<$~410~nm), photons are mostly absorbed and then re-emitted at longer wavelengths, which maximizes the detection efficiency of the PMTs. At long wavelengths ($>$~410~nm), photons mainly undergo Rayleigh scattering. A more detailed study of the wavelength-dependent absorption and re-emission can be found in Ref.~\cite{lsopmdl}. Additionally, the refractive indices at 420~nm are 1.50 and 1.34 for the LS and water, respectively. The difference in the refractive indices results in refraction and total reflection at the boundary of the two media, which affects the time-of-flight of the photons. When using the time information in the reconstruction, the time-of-flight is crucial, and it is calculated using the equation
	\begin{equation}
		 \mathrm {tof} = \sum_{m}\frac{d_{m}}{v_{m}},
		 \label{equ:tof}
	\end{equation} 
where $\mathrm {tof}$, $d_{m}$, and $v_{m}$ are the time-of-flight, optical path length, and effective light speed, respectively, and $m$ represents different media, in this case the LS and water, in the JUNO experiment. The acrylic sphere (thickness of 12~cm) and acrylic cover (thickness of 1~cm) in front of each PMT were ignored in this study because their refractive indices were similar to that of the LS and their thickness was small compared to that of the LS and water. 
	
\subsection{Optical path length}
	\begin{figure}[!ht]
		\centering
		\includegraphics[width=0.5\textwidth]{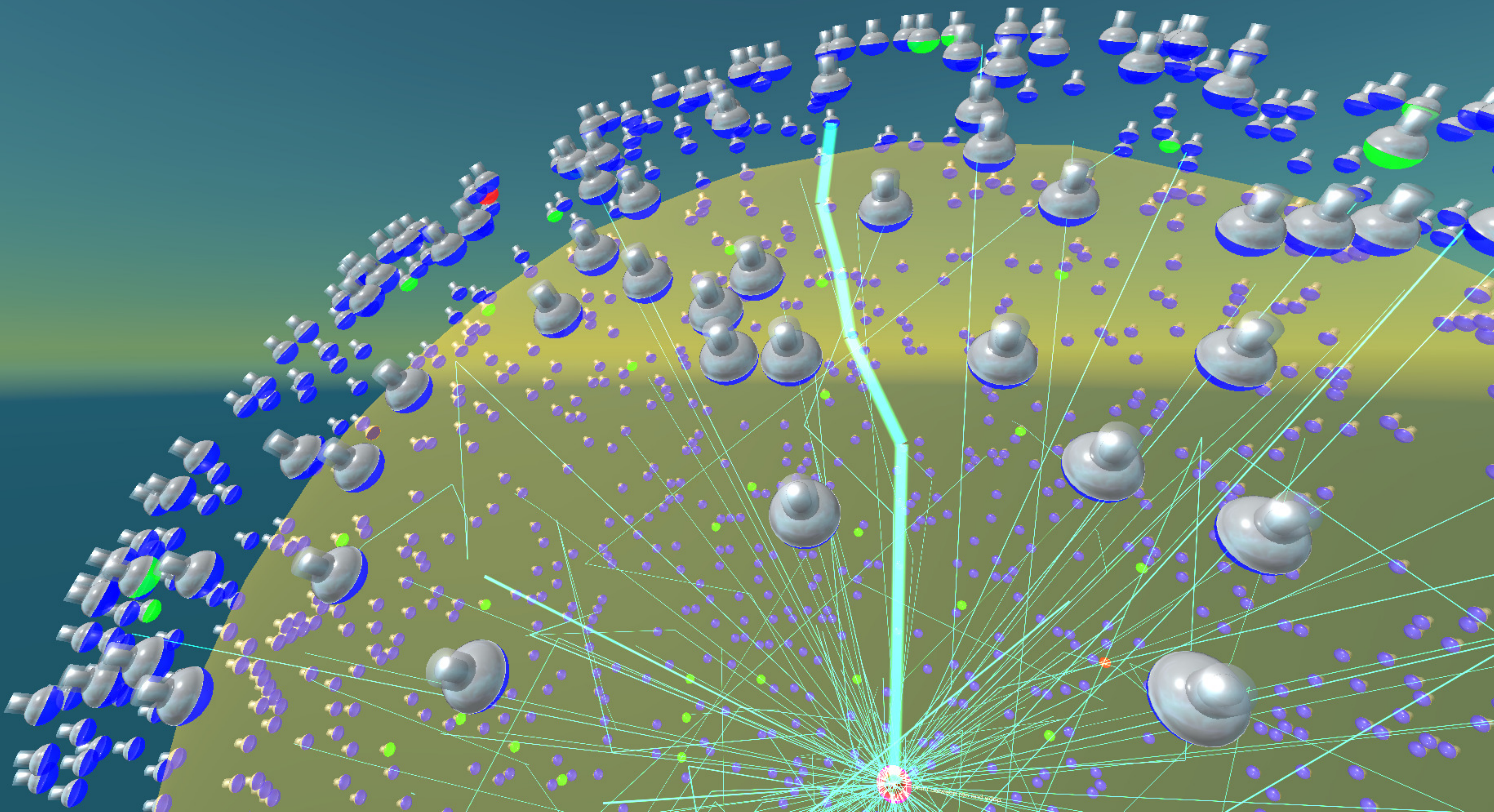}{\centering}									
		\caption{Event display of the optical path from the event vertex to the PMT in the JUNO simulation. The red circle ring is the event vertex and the gray bulbs with blue caps represent the PMTs.}
		\label{fig:path}			
	\end{figure}  
	
	The optical path length can be characterized by the start and end positions in the detector, which are the vertex of the event and the position of the PMTs, respectively. The bold cyan curve in Fig.~\ref{fig:path} shows a typical example of the optical path of photons detected by the PMT in the JUNO simulation using the event display~\cite{eventdisplay1,eventdisplay2}, and further examples are shown by the thin green curves. There are multiple physically possible paths between these two positions, each of which has a different optical path length, as follows: 
	\begin{enumerate}
		\item[$*$] owing to absorption and re-emission, the re-emitted photon is not in the same absorption position and the propagation direction also changes;
		\item[$*$] owing to scattering, the photon changes the original direction of the propagation;	 and	
		\item[$*$] owing to refraction and total reflection, the photon does not travel in a straight line. 
	\end{enumerate}

\begin{figure}[!ht]
		\centering
		\includegraphics[width=0.5\textwidth]{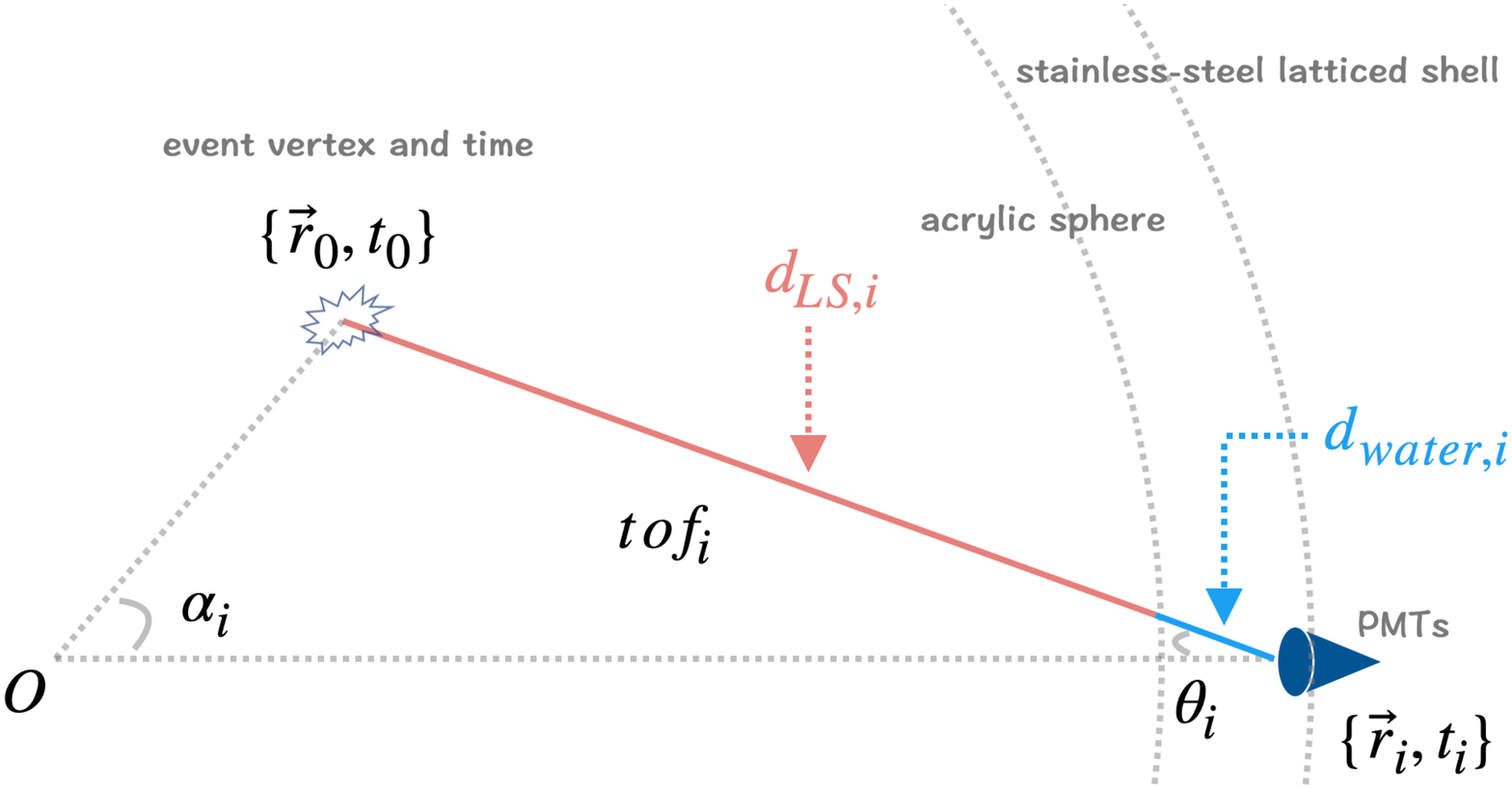}{\centering}									
		\caption{Optical path length from the event vertex to the $i$th PMT. $O$ denotes the center of the detector.}
		\label{fig:lightpath}			
	\end{figure}

	As shown in Fig.~\ref{fig:path}, owing to the various aforementioned optical processes, it is difficult to predict the actual optical path length for each photon. In this paper, a simple optical model is proposed, which uses a straight line connecting the vertex and the PMTs to calculate the optical path length (Fig.~\ref{fig:lightpath}), and combines with the effective light speed to correct for the time-of-flight. Using this simple optical model reasonable results can be obtained, as discussed in Sec.~\ref{sec:tlh}.

	In Fig.~\ref{fig:lightpath}, \{$\vec{r}_{0},t_{0}$\} represents the event vertex and start time, \{$\vec{r}_{i},t_{i}$\} is the position of the $i$th PMT and the time of the earliest arriving photon detected by it. The angle between the normal direction of the $i$th PMT and the vector of the position of the $i$th PMT pointing to the event vertex is $\theta_{i}$ and $\alpha_{i}=\arccos(\hat{\vec{r}}_{0}\cdot\hat{\vec{r}}_{i})$. The optical path length of the photon arriving at the $i$th PMT is $d_{pathlength,i} = |\vec{r}_{i} - \vec{r}_{0}| = d_{LS,i} + d_{water,i}$ and the corresponding time-of-flight is $\mathrm {tof}_{i}$. The optical path length in the LS and water can be calculated by simply solving the trigonometric equation.

\subsection{Effective light speed}	
	According to Ref.~\cite{lsopmdl}, the emission spectrum of scintillation photons is in the range of approximately 300––600 nm. Typically, the group velocity of the wave packet is used to describe the photon propagation in the medium, which is given by the equation
	\begin{equation}
		v_{g}(\lambda) = \frac{c}{n(\lambda)-\lambda\frac{\partial{n(\lambda)}}{\partial\lambda}},
		\label{equ:vg}
	\end{equation} 		
	where $v_{g}$ is the group velocity, $c$ is the speed of light in vacuum, $n$ is the refractive index, and $\lambda$ is the wavelength.

\begin{figure}[!ht]
		\centering
		\includegraphics[width=0.5\textwidth]{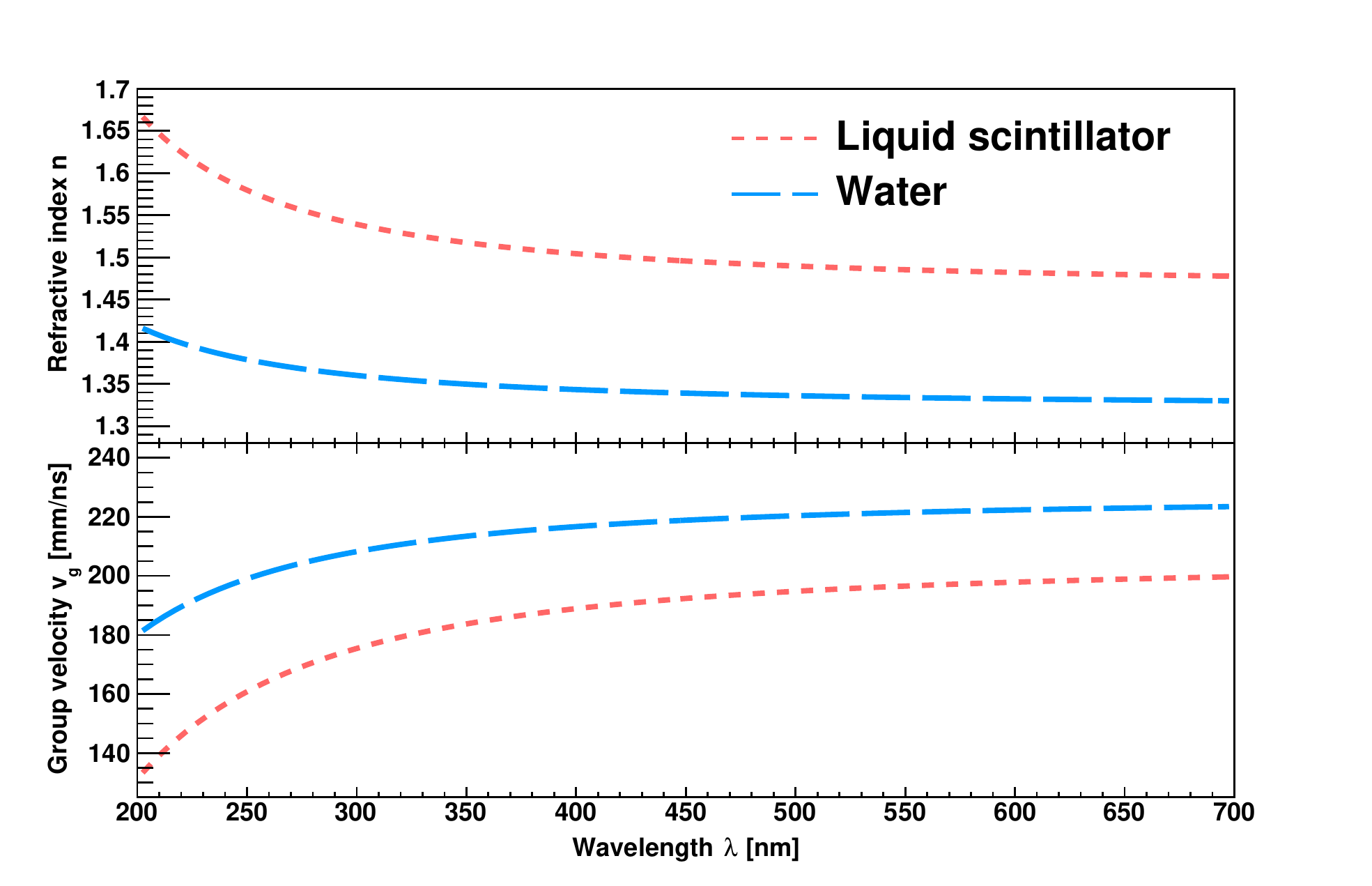}{\centering}									
		\caption{Dependence of the refractive index (upper panel) and group velocity (lower panel) on the wavelength in the LS and water.}
		\label{fig:groupvelocity}			
	\end{figure}

	By fitting the Sellmeier equation~\cite{optics}, which describes the dispersion of the measurement in Refs.~\cite{rayleighscat} and~\cite{h2oindex}, the refractive index of the LS and water at different wavelengths is shown in the upper panel of Fig.~\ref{fig:groupvelocity}. The group velocity of the LS and water can be calculated using Eq.~\ref{equ:vg} at different wavelengths, as shown in the lower panel of Fig.~\ref{fig:groupvelocity}.
	
	The propagation speed of photons in water ($v_{water}$) was determined as the average speed weighted by the probability density function of the photon wavelength, which was obtained from a Monte Carlo (MC) simulation. As the absorption and re-emission change the initial wavelength, determining the propagation speed of photons in the LS ($v_{LS}$) is more complicated. To consider all wavelength-dependent effects that affect the propagation speed of photons, the effective light speed $v_{eff}$ is introduced. In addition, $v_{eff}$ also mitigates the effects by the simplified optical model, which, for example, ignores the refraction at the interface between the LS and water, as well as the change in the optical path length due to Rayleigh scattering. The exact value for $v_{eff}$ can be determined using a data-driven method based on the calibration data as follows: place $\gamma$ sources along the Z-axis, use $v_{LS}$ at 420~nm as the initial value of $v_{eff}$ in the reconstruction algorithm and then, calibrate $v_{eff}$ such that the source positions can be appropriately reconstructed. As no calibration data was available for JUNO, in this study, simulated calibration data were used, and the optimized values for the effective refractive index (c/$v_{eff}$) were 1.546 in the LS and 1.373 in water. In the future, the same method can be applied to the experimental calibration data.
	
\section{Initial value for vertex and time}
\label{sec:initvalue}
	The TMinuit package~\cite{tminuit} was used for the minimization procedure in the time likelihood and in the charge likelihood algorithm introduced in Secs.~\ref{sec:tlh} and~\ref{sec:clh}. When there are multiple local minima in the parameter space, an inaccurate initial value results in local instead of global minima, resulting in a lower reconstruction efficiency. For detectors such as JUNO, the initial value needs to be treated carefully because of the total reflection, as discussed in the following subsections.	
	
\subsection{Charge-based algorithm}
	The charge-based algorithm is essentially based on the charge-weighted average of the positions of the PMTs in an event, and the event vertex can be determined using the equation
	\begin{equation}
		 \vec{r}_{0} = a\cdot\frac{\sum_{i}{q_{i}\cdot\vec{r}_{i}}}{\sum_{i}{q_{i}}},
	\end{equation} 
	where $q_{i}$ is the charge of the pulses detected by the $i$th PMT and $\vec{r}_{0}$ and $\vec{r}_{i}$ are defined in Fig.~\ref{fig:lightpath}. A scale factor $a$ is introduced because the charge-based algorithm is inherently biased and an ideal point-like event in a spherical detector is covered by a uniform photocathode. Even if all propagation-related effects, such as absorption and scattering are ignored, the result of a simple integral of the intersections of all photons with the sphere surface shows that the reconstructed position of the event is 2/3 of the true position. The value of $a$ can be tuned based on the calibration data along the $Z$-axis. In this study, $a=1.3$ was used, which was sufficient to provide an initial estimate for the event vertex. 

	\begin{figure}[!ht]
		\centering
		\includegraphics[width=0.5\textwidth]{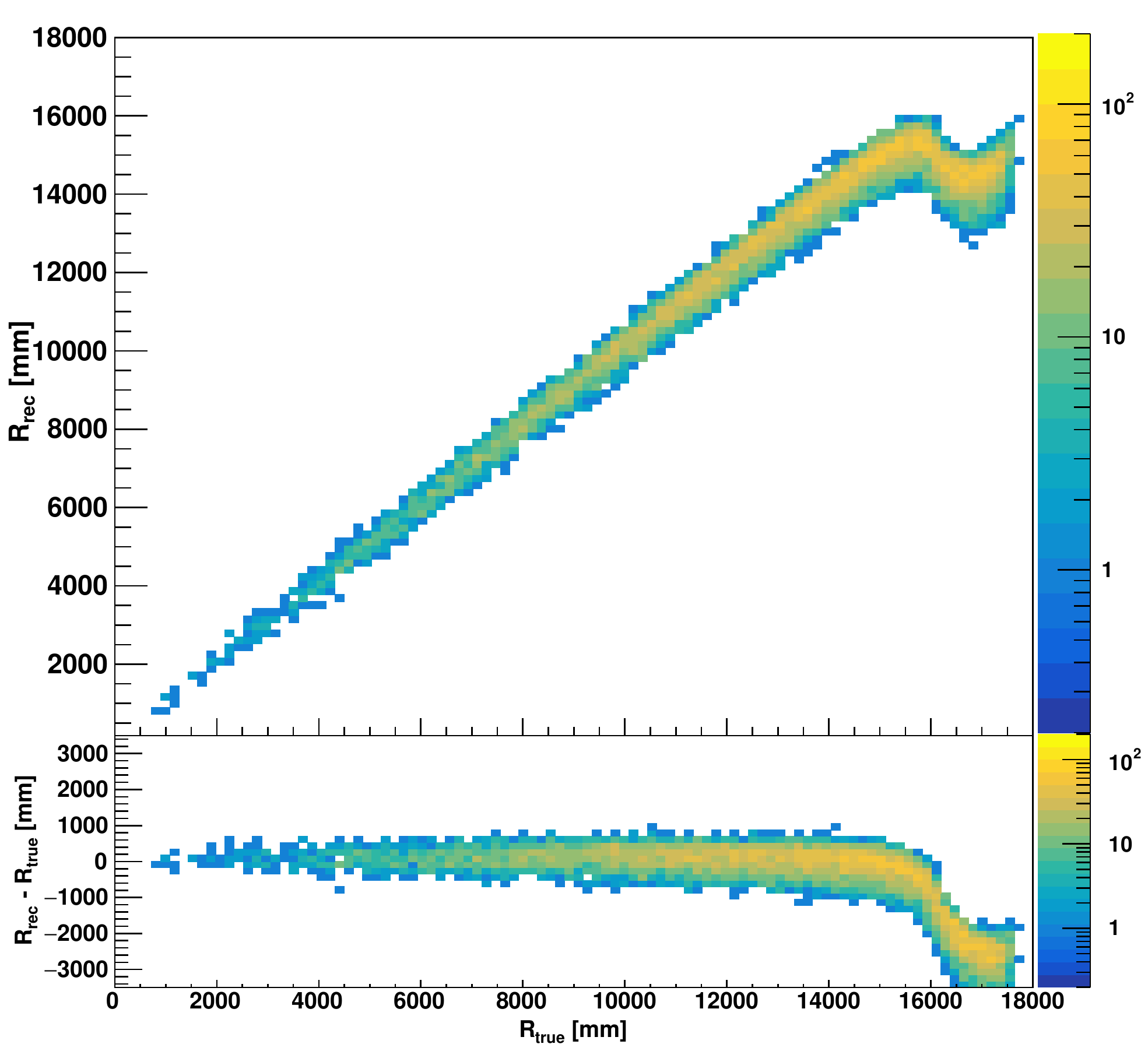}{\centering}		
		\caption{Heatmap of $R_{rec}$ (upper panel) and $R_{rec}-R_{true}$ (lower panel) as a function of $R_{true}$ for 4-MeV $e^{+}$ uniformly distributed in space calculated by the charge-based algorithm.}
		\label{fig:cb_edepr_recr_4mev}			
	\end{figure}   
	
	As can be seen in Fig.~\ref{fig:cb_edepr_recr_4mev}, even with the scale factor, owing to total reflection, the reconstructed vertex deviates up to 3~m near the detector boundary. According to Ref.~\cite{optics}, total reflection occurs only when the event vertex is located at an $R$ larger than $R_{LS}\cdot{n_{water}/n_{LS}}\approx15.9$~m, where $R_{LS}$ is the radius of the acrylic sphere, $n_{LS}$ and $n_{water}$ are the refractive indices in the LS and water, respectively. The total reflection region is defined as $R > 15.9$~m while $R < 15.9$~m is the central region. If the result from the charge-based algorithm is used as the initial value for the time likelihood algorithm, approximately 18\% of events is reconstructed at a local minimum position. In addition, it should be noted that the charge-based algorithm is not able to provide an initial value for the event generation time $t_{0}$. Therefore, a fast time-based algorithm needs to be introduced, which can provide more accurate initial values.	
	
\subsection{Time-based algorithm}
	The time-based algorithm uses the distribution of the time-of-flight correction time $\Delta{t}$ (defined in Eq.~\ref{equ:tofcorrti}) of an event to reconstruct its vertex and $t_{0}$. In practice, the algorithm finds the reconstructed vertex and $t_{0}$ using the following iterations: 

	\begin{enumerate}	
		\item Apply the charge-based algorithm to obtain the initial vertex.
		\item Calculate time-of-flight correction time $\Delta{t}$ for the $i$th PMT as
			\begin{equation}
				\Delta{t}_{i}(j) = t_{i} - \mathrm {tof}_{i}(j),
			\label{equ:tofcorrti}
			\end{equation} where $j$ is the iteration step and $t_{i}$, $\mathrm {tof}_{i}$ are defined in Fig.~\ref{fig:lightpath}.
			Plot the $\Delta{t}$ distribution for all triggered PMTs, and label the peak position as $\Delta{t}^{peak}$.
		\item Calculate the correction vector $\vec\delta[\vec{r}(j)]$ as
			\begin{equation}
				\vec\delta[\vec{r}(j)] =\frac{\sum_{i}(\frac{\Delta{t}_{i}(j) - \Delta{t}^{peak}(j)}{\mathrm {tof}_{i}(j)})\cdot(\vec{r}_{0}(j)-\vec{r}_{i})}{N^{peak}(j)},
			\end{equation} where $\vec{r}_{0}$, and $\vec{r}_{i}$ are defined in Fig.~\ref{fig:lightpath}.	
			To minimize the effect of scattering, reflection, and dark noise on the bias of the reconstructed vertex, only the pulses appearing in the $(-10~\rm ns, +5~\rm ns)$ window around $\Delta{t}^{peak}$ are included. The time cut also suppresses the effect of the late scintillation photons. The number of triggered PMTs in the window is $N^{peak}$.
		\item If $\vec\delta[\vec{r}(j)] < 1~\rm mm$ or $j=100$, stop the iteration; otherwise, update the vertex with $\vec{r}_{0}(j+1) = \vec{r}_{0}(j) + \vec\delta[\vec{r}(j)]$ and go to step 2 to start a new round of iteration.
	\end{enumerate}

\begin{figure}[!ht]
		\centering
		\includegraphics[width=0.5\textwidth]{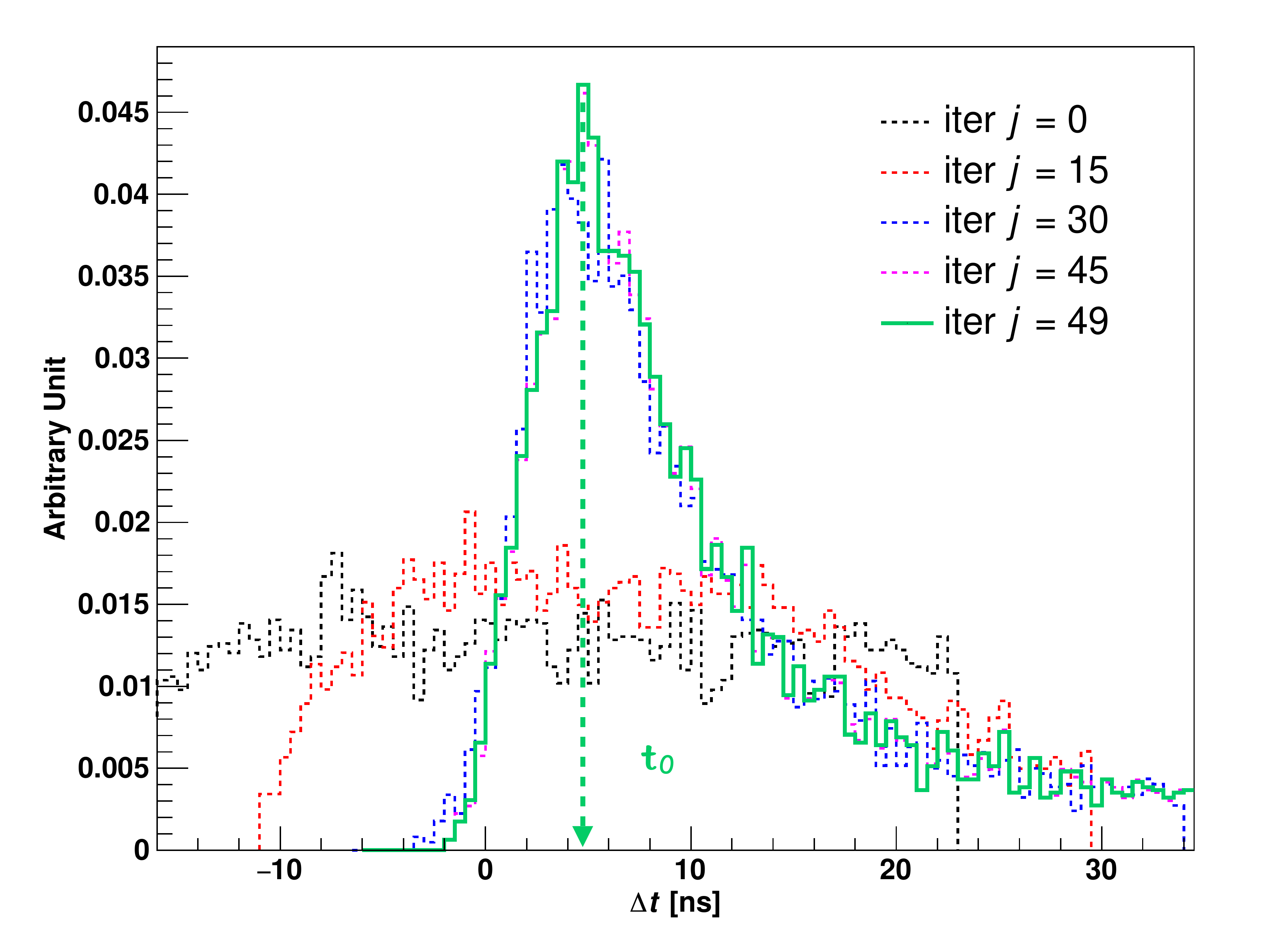}{\centering}
		\caption{$\Delta{t}$ distribution at different iteration steps $j$.}
		\label{fig:tb_illus_iter}			
	\end{figure}

	The distribution of $\Delta{t}$ at different iteration steps is shown in Fig.~\ref{fig:tb_illus_iter}. At the beginning of the iteration, the $\Delta{t}$ distribution is wide because the initial vertex is far from the true vertex. As the number of iterations increases, the $\Delta{t}$ distribution becomes more concentrated. Finally, when the requirement in step 4 is satisfied, the iteration stops. In the final step, $\vec{r}_{0}$ is the reconstructed vertex and $\Delta{t}^{peak}$ is the reconstructed time $t_{0}$. 		
			
	After the time-of-flight correction, the $\Delta{t}$ distribution is independent of the event vertex. However, because the earliest arrival time is used, according to the first-order statistic, as discussed in Ref.~\cite{fos,fos2,fos3}, $t_{i}$ is related to the number of photoelectrons $N^{i}_{pe}$ detected by $i$th PMT. To reduce the bias of the vertex reconstruction, the following form of the time--$N_{pe}$ correction is applied, and in Eq.~\ref{equ:tofcorrti} $t_{i}$ is replaced by $t'_{i}$:
	\begin{equation}
		t'_{i}= t_{i} - p0/\sqrt{N^{i}_{pe}} - p1 - p2/N^{i}_{pe}.
	\end{equation}

	The parameters $(p0,p1,p2)$ with the corresponding values of (9.42, 0.74, $-$4.60) for Hamamatsu PMTs and (41.31, $-$12.04, $-$20.02) for NNVT PMTs were found to minimize the bias and energy dependence of the reconstruction in this study. The difference in the parameters is mainly due to the difference in the TTSs of the PMTs. Following the correction, the times of different PMTs with different values of $N_{pe}$ are aligned.

	\begin{figure}[!ht]
		\centering
		\includegraphics[width=0.5\textwidth]{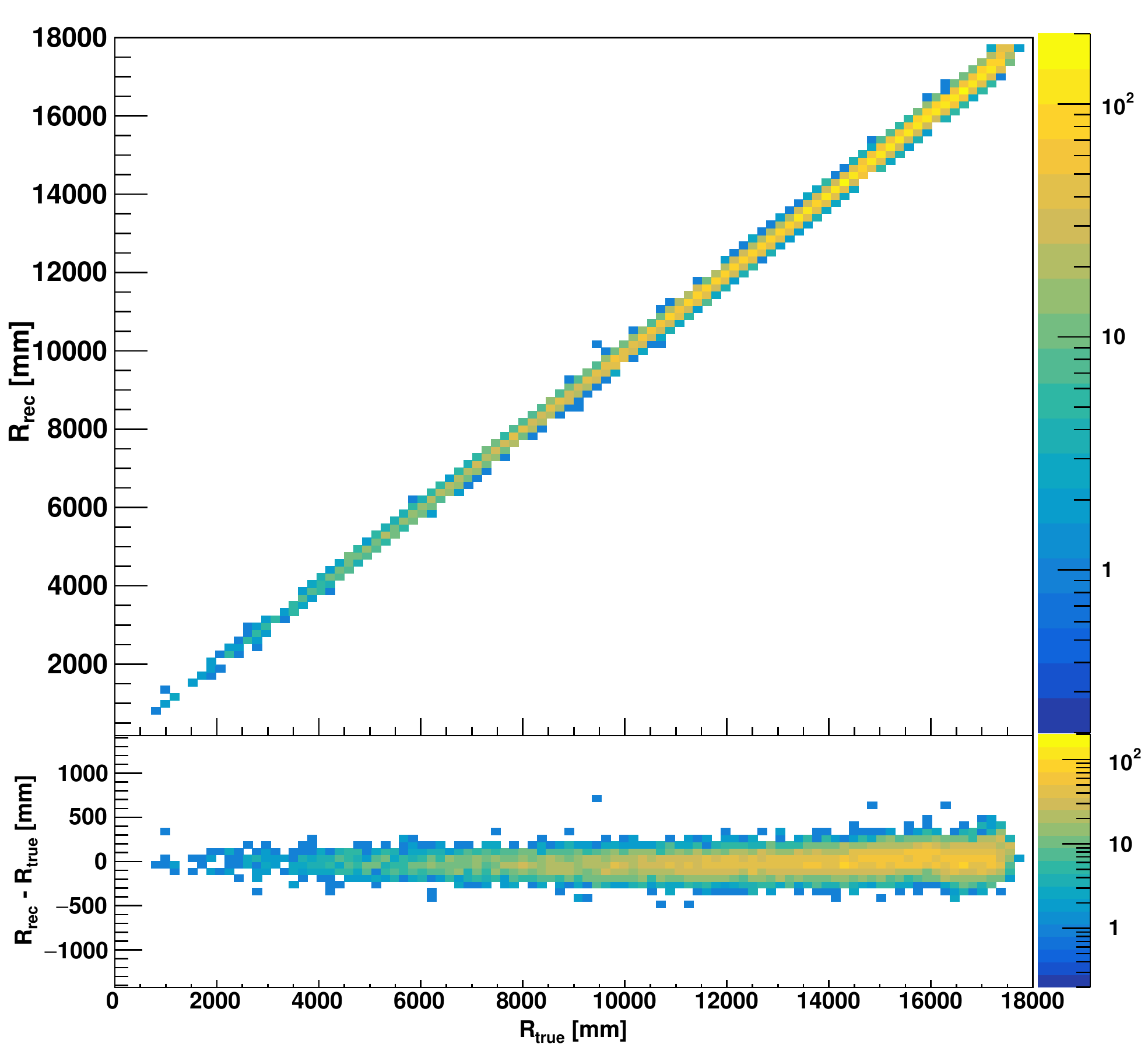}{\centering}		
		\caption{Heatmap of $R_{rec}$ (upper panel) and $R_{rec}-R_{true}$ (lower panel) as a function of $R_{true}$ for 4-MeV $e^{+}$ uniformly distributed in space calculated by the time-based algorithm.}
		\label{fig:tb_edepr_recr_4mev}			
	\end{figure}   		
	
	As shown in Fig.~\ref{fig:tb_edepr_recr_4mev}, the time-based algorithm provided a more accurate reconstructed vertex than the charge-based algorithm (Fig.~\ref{fig:cb_edepr_recr_4mev}). In addition, after the time--$N_{pe}$ correction, the reconstruction shows no obvious bias within the entire detector, even in the total reflection region. The reconstructed result was used as the initial value for the time likelihood algorithm.	

\section{Time likelihood algorithm}
\label{sec:tlh}
\subsection{Principle of the algorithm }	
	The time likelihood algorithm uses the scintillator response function to reconstruct the event vertex. The variable residual time $t_{res}(\vec{r}_{0},t_{0})$ for the $i$th PMT can be described as
	\begin{equation}
		t^{i}_{res}(\vec{r}_{0},t_{0}) = t_{i} - \mathrm {tof}_{i} - t_{0},
	\end{equation} 	
	where $t^{i}_{res}$ is the residual time of the $i$th PMT and $\vec{r}_{0}$, $t_{0}$, $t_{i}$, and $\mathrm {tof}_{i}$ are defined in Fig.~\ref{fig:lightpath}.

	The scintillator response function mainly consists of the emission time profile of the scintillation photons and the TTS and the dark noise of PMTs. In principle, the additional delays introduced by the absorption, re-emission, scattering, and total reflection of the photon arriving to the PMT depend on the distance between the emission position and the individual PMTs. However, the differences are only noticeable for the late arrival hits, which are largely suppressed by the requirement for the earliest arriving photons in the time likelihood algorithm. Therefore, in the first-order approximation, the scintillator response function can be considered to be the same for all positions inside the scintillator. The scintillator response function can be described as follows.
	
	As described in Sec.~\ref{sec:op}, when a charged particle interacts with a scintillator molecule, the molecule is excited, then de-excites, and emits photons. Typically, the scintillator has more than one component; thus, the emission time profile of the scintillation photons, $f(t_{res})$, can be described as 

	\begin{equation}
		f(t_{res}) = \sum_{k}\frac{\rho_{k}}{\tau_{k}}e^{\frac{-t_{res}}{\tau_{k}}}, \sum_{k}\rho_{k} = 1,
	\end{equation} 	
where each ${k}$ component is characterized by its decay time $\tau_{k}$ and intensity $\rho_{k}$. The different components result from the different excited states of the scintillator molecules.
		
	To consider the spread in the arrival time of photons at the PMTs, a convolution with a Gaussian function is applied, given by
	\begin{equation}
		g(t_{res}) = \frac{1}{\sqrt{2\pi}\sigma}e^{-\frac{(t_{res}-\nu)^2}{2\sigma^2}}\cdot{f(t_{res})}.
	\end{equation} 	where $\sigma$ is the TTS of PMTs and $\nu$ is the average transit time.

	The dark noise, which occurs without incident photons in the PMTs, is not correlated with any physical event. The fraction of the dark noise in the total number of photoelectrons $\varepsilon_{dn}$ can be calculated based on the data acquisition (DAQ) windows, dark noise rate, and light yield of the LS. The probability of dark noise $\varepsilon(t_{res})$ is constant over time, where $\int_{DAQ}{\varepsilon(t_{res})dt_{res}}=\varepsilon_{dn}$. By adding $\varepsilon(t_{res})$ to $g(t_{res})$ and renormalizing its integral to 1, the probability density function (PDF) of the scintillator response function can be written as
	
	\begin{equation}
		p(t_{res}) = (1-\varepsilon_{dn} )\cdot{g(t_{res})} +\varepsilon(t_{res}).
		\label{equ:pdfwttsdn}
	\end{equation}	
	
	The distribution of the residual time $t_{res}$ of an event for a hypothetical vertex can be compared with $p(t_{res})$. The best-fitting vertex and $t_{0}$ are chosen by minimizing the negative log-likelihood function
	\begin{equation}
		\mathcal{L} {(\vec{r}_{0},t_{0})} = -\ln(\prod_{i}p(t^{i}_{res})).
	\end{equation}	

	The parameters in Eq.~\ref{equ:pdfwttsdn} can be measured experimentally~\cite{tres1,tres2,tres3,tres4}. In this work, the PDF from the MC simulation for the methodology study was employed.
		
\subsection{Probability density function}	
	\begin{figure}[!ht]
		\centering
		\includegraphics[width=0.5\textwidth]{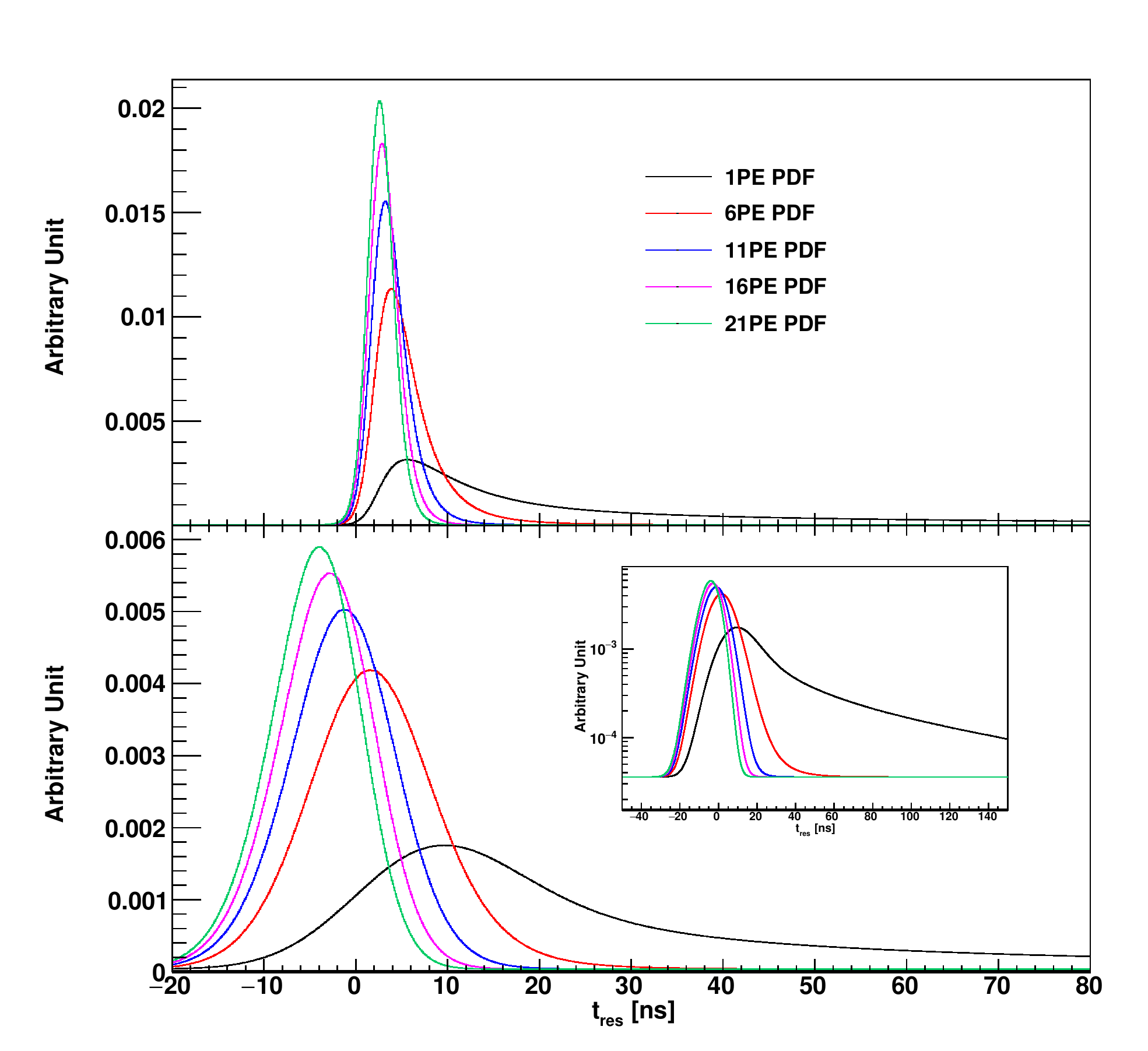}{\centering}		
		\caption{PDF of the scintillator response function for PMTs detecting different numbers of photoelectrons. The upper panel shows the response function for Hamamatsu, the lower panel for the NNVT PMTs.}
		\label{fig:tlh_pdf}			
	\end{figure}   
	
	\begin{figure*}[!ht]
		\centering
		\subfigure{\includegraphics[width=0.33\textwidth]{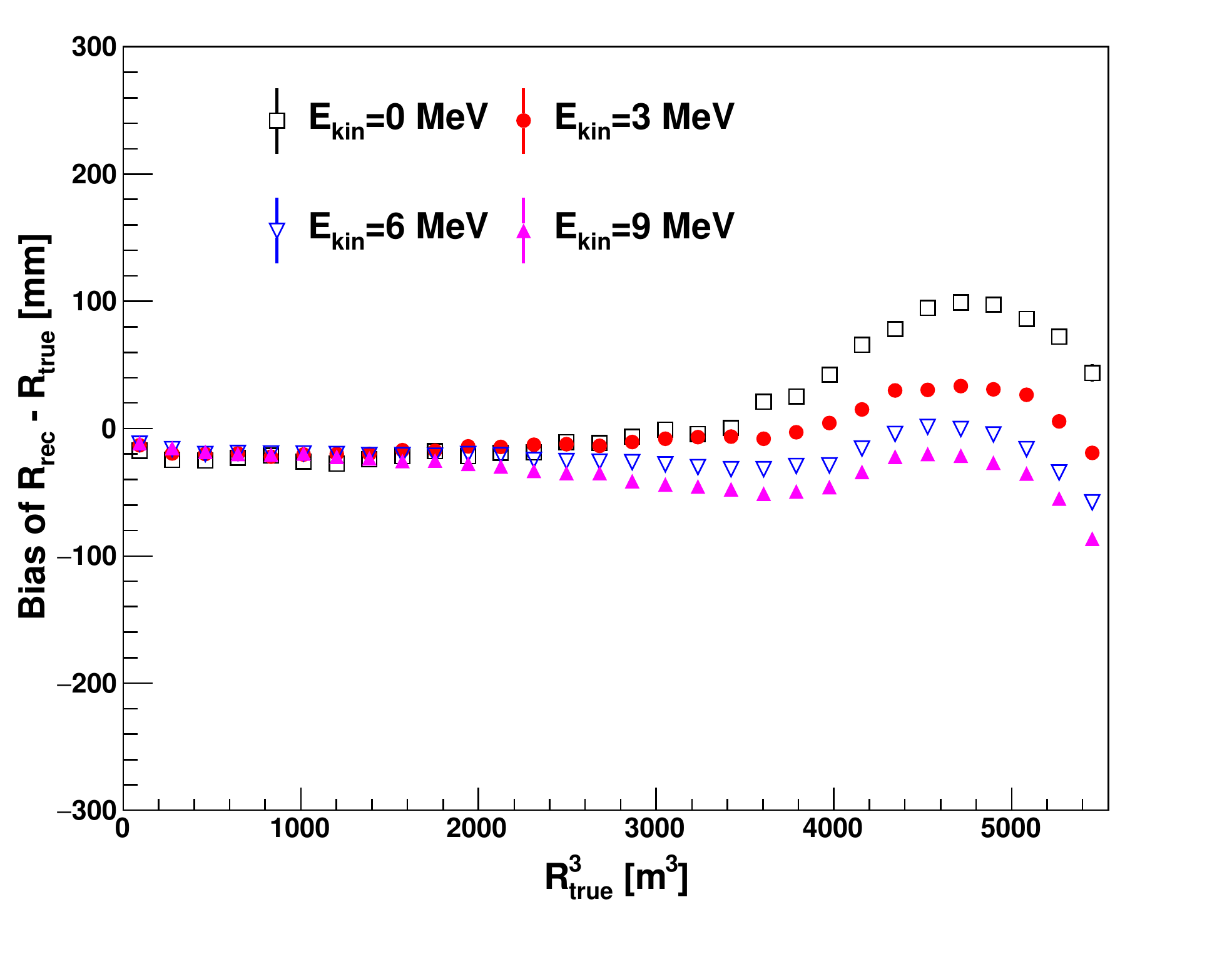}}
		\subfigure{\includegraphics[width=0.33\textwidth]{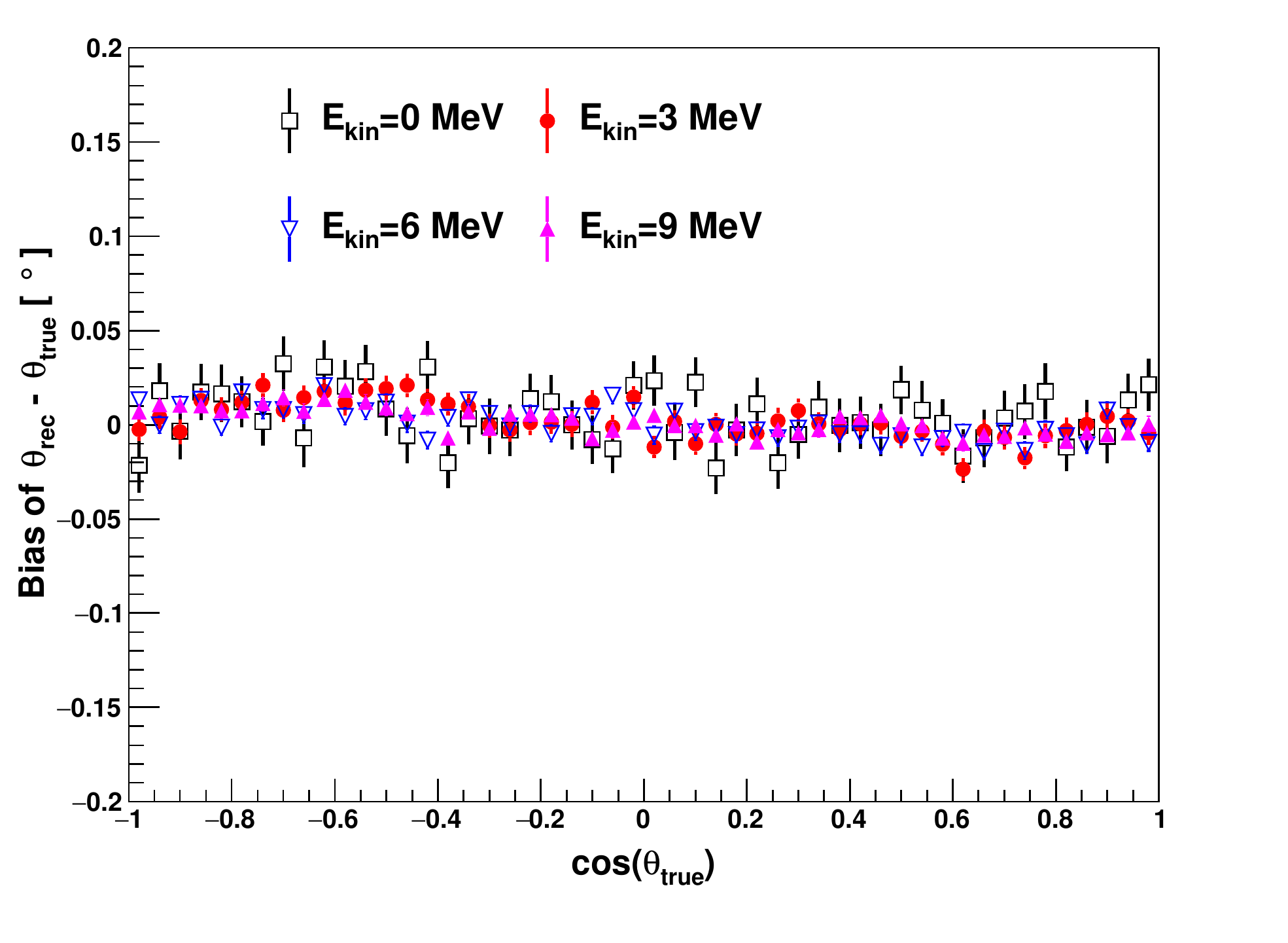}}
		\subfigure{\includegraphics[width=0.33\textwidth]{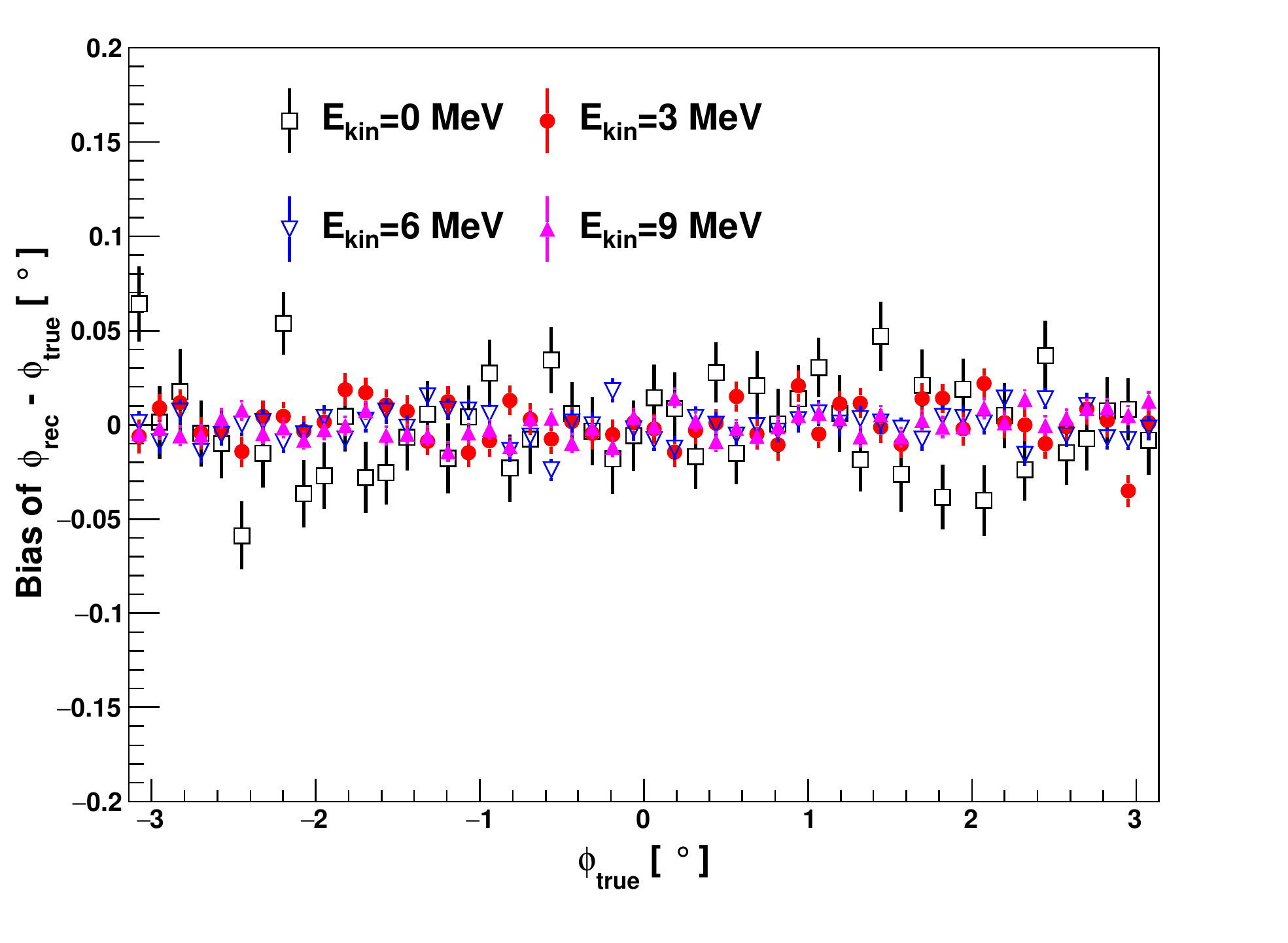}}
		\caption{Bias of the reconstructed $R$ (left panel), $\theta$ (middle panel), and $\phi$ (right panel) for different energies calculated by the time likelihood algorithm.}
		\label{fig:tlh_bias}			
	\end{figure*}

	\begin{figure*}[!ht]
		\centering
		\subfigure{\includegraphics[width=0.33\textwidth]{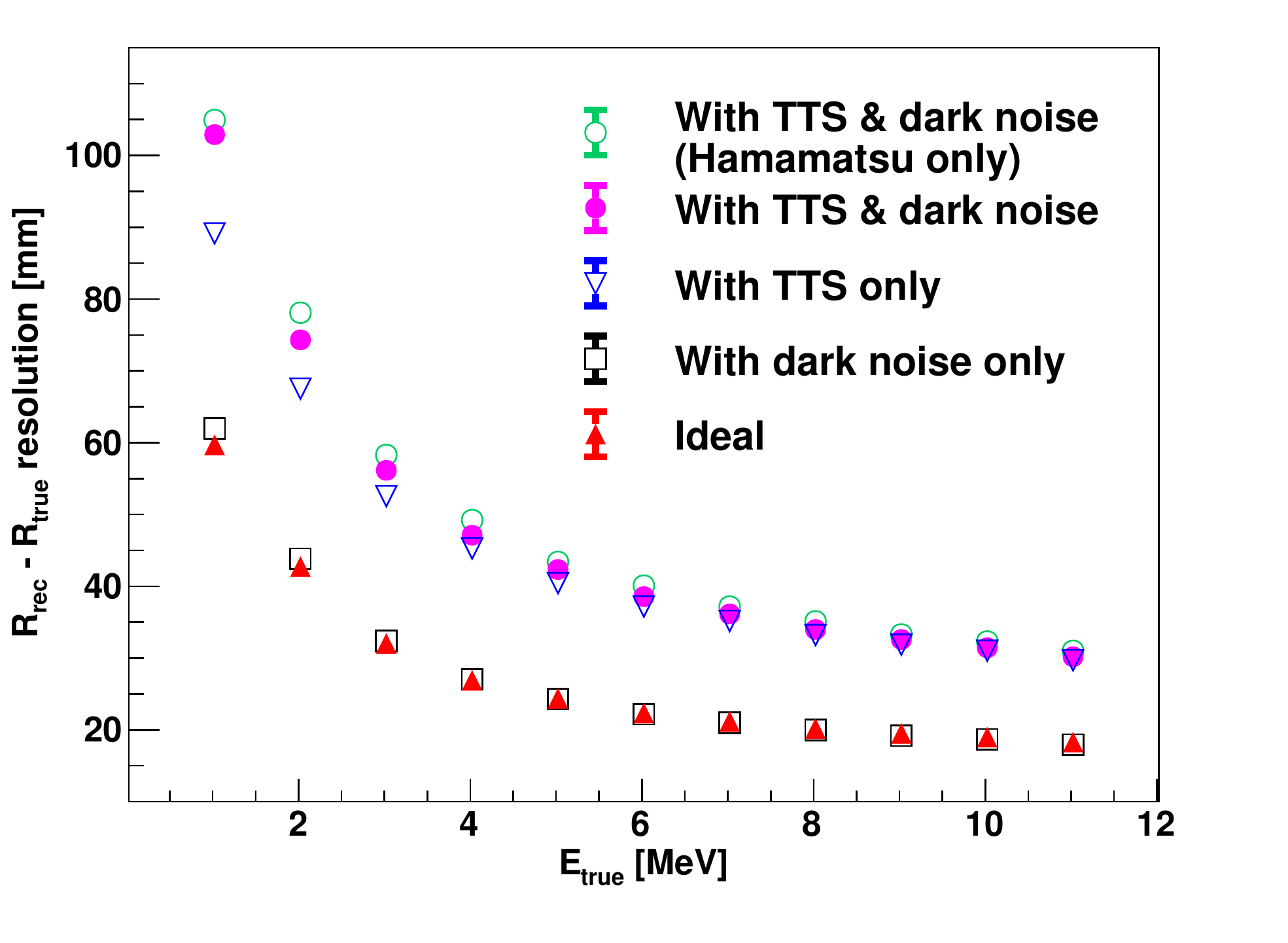}}
		\subfigure{\includegraphics[width=0.33\textwidth]{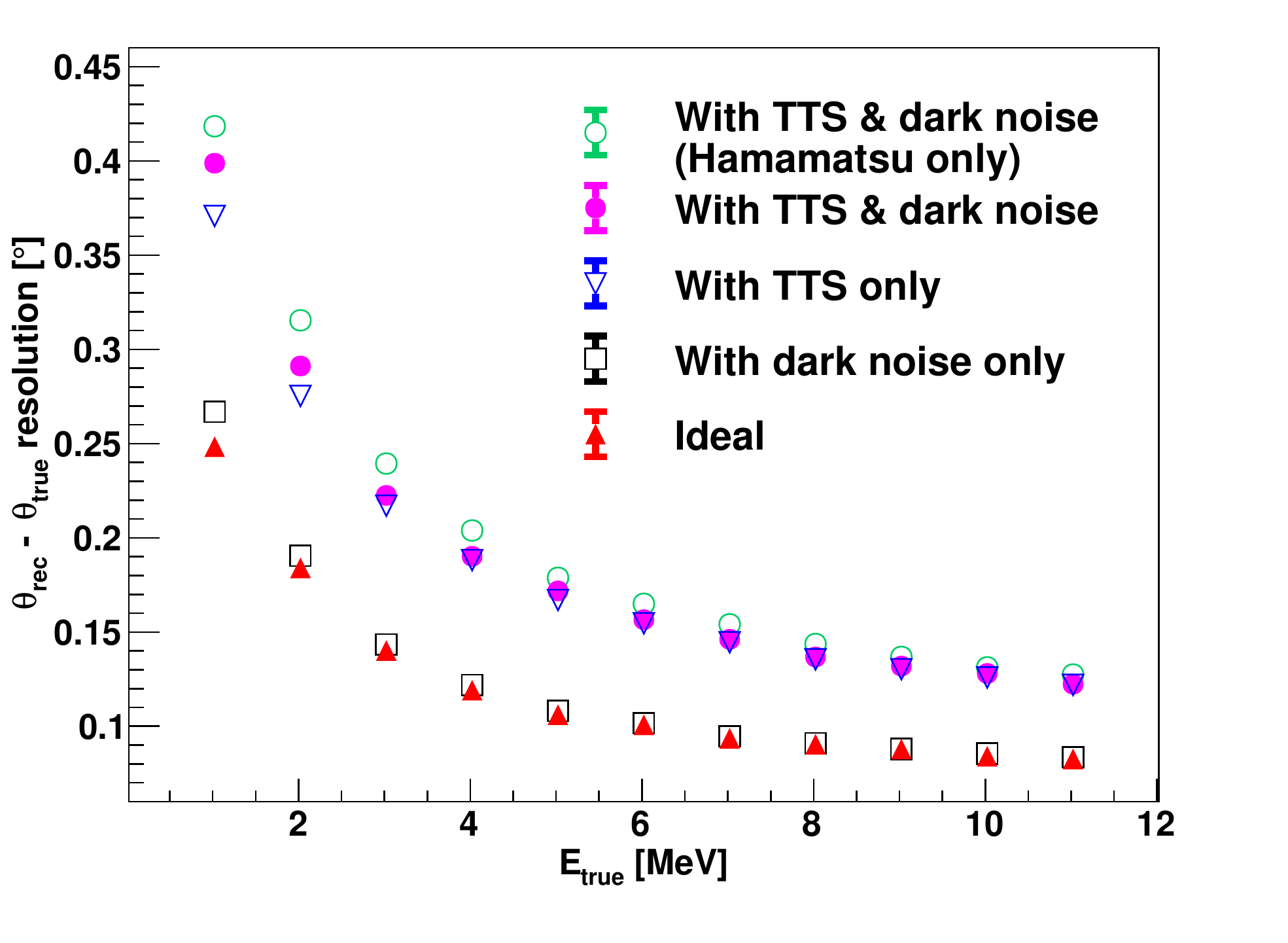}}
		\subfigure{\includegraphics[width=0.33\textwidth]{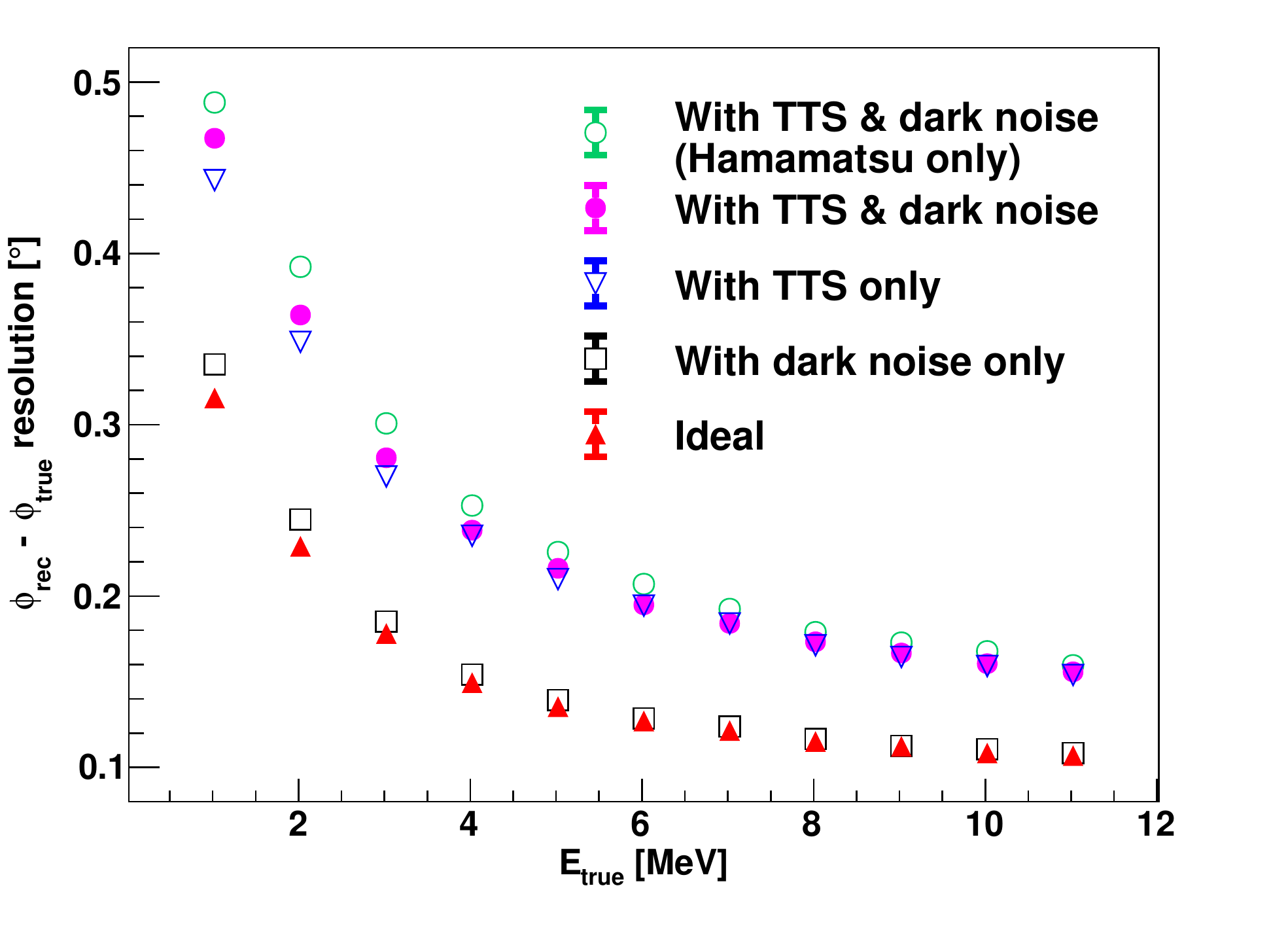}}
		\caption{Resolution of the reconstructed $R$ (left panel), $\theta$ (middle panel), and $\phi$ (right panel) as a function of energy calculated by the time likelihood algorithm. Different colors represent different PMT configurations.}	
		\label{fig:tlh_res}			
	\end{figure*}
		
	The PDF of the scintillator response function for PMTs detecting a single photoelectron was obtained from the MC simulation, using a 4.4-MeV $\gamma$ source located at the center of the detector, such that the distance to all PMTs is the same. For PMTs detecting multiple photoelectrons, the time of the earliest arriving photon is biased toward an earlier time. Therefore, the PDF need to be modified according to the first-order statistic of $p(t_{res})$ or the so-called first photoelectron timing technique~\cite{fos,fos2,fos3} as
	
	\begin{equation}
		p_{N_{pe}}(t_{res}) = N_{pe}p(t_{res})(\int_{t_{res}}^{\infty}p(x)dx)^{N_{pe}-1},
	\end{equation}	
	where $p_{N_{pe}}(t_{res})$ is the PDF of the scintillator response function when the PMTs detect $N_{pe}$ hits.

	The PDF of two kinds of PMTs is shown in Fig.~\ref{fig:tlh_pdf}: the upper panel is for Hamamatsu while the lower panel is for NNVT PMTs. As the PDF is affected by the time resolution of the PMTs, the PDF of the NNVT is wider because of its inferior TTS. The inset in the lower panel shows the PDF on a logarithmic scale, and the time constant contribution of the dark noise $\varepsilon(t_{res})$ is clearly visible.

\subsection{Reconstruction performance}
	The reconstructed vertex was compared with the true vertex in spherical coordinates ($R, \theta, \phi$) for the MC $e^{+}$ samples and fitted with a Gaussian function to analyze the bias and resolution. The bias of the reconstruction is shown in Fig.~\ref{fig:tlh_bias}, where different colors represent events with different energies. As can be seen in the left panel of Fig.~\ref{fig:tlh_bias}, the reconstructed $R$ is consistent with the true value in the central region, while an energy-dependent bias behavior is noticeable near the detector boundary. Given its regular bias behavior, the bias can be corrected with an energy-dependent correction. Moreover, although the reconstructed $R$ is biased, there is no bias in $\theta$ and $\phi$, as shown in the middle and right panels of Fig.~\ref{fig:tlh_bias}, respectively.
	
	The spatial resolution of the vertex reconstruction as a function of energy is shown in Fig.~\ref{fig:tlh_res}. The $R$ bias was corrected before the analysis of the resolution. To study the individual effect of the TTS and dark noise on the vertex reconstruction, different MC samples were produced with and without these effects. The vertex reconstruction results are shown in Fig.~\ref{fig:tlh_res}. The magenta circles represent the default PMT configuration, as described in Sec.~\ref{sec:junodetector}. The red triangles represent an ideal configuration, which assumes perfect PMTs without the effects of the TTS and dark noise. The black squares represent the configuration of PMTs including only the dark noise effect, while the blue inverted triangles represent the PMT configuration including only the TTS effect. The exact values of the vertex resolution at 1.022~MeV and 10.022~MeV are summarized in Tables~\ref{tab:tlhres_1mev} and~\ref{tab:tlhres_10mev}, respectively. The energy $E_{true}$ includes the energy of the annihilation gamma rays. The light yield was approximately 1300 detected $N_{pe}$ per 1~MeV of deposited energy in JUNO, and the energy nonlinearities on the light yield were ignored in the approximation. As can be seen in Tables~\ref{tab:tlhres_1mev} and~\ref{tab:tlhres_10mev}, the dark noise has no effect at high energy and its effect at low energy is also highly limited. The largest effect results from the TTS in the time likelihood algorithm. The energy-dependent vertex resolution is approximately proportional to $1/\sqrt{N_{pe}}$~\cite{fos2}.
	
	\begin{table}[!htb]
    		\caption{Vertex resolution for different PMT configurations at 1.022~MeV (detection of $\sim$1328 $N_{pe}$ in total, corresponding to $\sim$370 $N_{pe}$ detected by Hamamatsu PMTs).}
		\begin{tabular*}{8cm} {@{\extracolsep{\fill} } cccc}
 	       		\toprule
        		 	PMT configuration 	& $R$ (mm)	& $\theta$	 (degrees)		& $\phi$ (degrees) \\
			\midrule
        			Ideal 		& 60 			& 0.25				& 0.31 \\
        			With dark noise only		& 62	& 0.27 				& 0.34 \\
        			With TTS only		 	& 89 	& 0.37 				& 0.44 \\
        			With TTS and dark noise 	& 103	& 0.40 				& 0.47 \\	
        			\makecell{With TTS and dark noise\\(Hamamatsu PMTs only)}	  	& 105		& 0.42 		& 0.49 \\						
        			\bottomrule
		\end{tabular*}
    		\label{tab:tlhres_1mev}
	\end{table}	
	
	\begin{table}[!htb]
    		\caption{Vertex resolution for different PMT configurations at 10.022~MeV (detection of $\sim$13280 $N_{pe}$ in total, corresponding to $\sim$ 3700 $N_{pe}$ detected by Hamamatsu PMTs).}
		\begin{tabular*}{8cm} {@{\extracolsep{\fill} } cccc}
 	       		\toprule
        		 	PMT configuration 	& $R$ (mm)	& $\theta$	 (degrees)		& $\phi$ (degrees) \\
			\midrule
        			Ideal 				& 19 		& 0.08				& 0.11 \\
        			With dark noise only		& 19		& 0.08 				& 0.11 \\
        			With TTS only		 	& 31		& 0.13 				& 0.16 \\
        			With TTS and dark noise 	& 31		& 0.13				& 0.16 \\	
        			\makecell{With TTS and dark noise\\(Hamamatsu PMTs only)}	  	& 32		& 0.14 		& 0.17 \\							
        			\bottomrule
		\end{tabular*}
    		\label{tab:tlhres_10mev}
	\end{table}

	Owing to the low time resolution of the NNVT PMTs, in Fig.~\ref{fig:tlh_res} only the reconstruction using Hamamatsu PMTs is shown (green circles). In this study, we found that the vertex resolution with Hamamatsu PMTs was similar to that of using all PMTs. The reconstruction speed was 3.5 times faster, because the fraction of the Hamamatsu PMTs was approximately 28\% of all PMTs in the CD.

	\begin{figure}[!ht]
		\centering
		\includegraphics[width=0.5\textwidth]{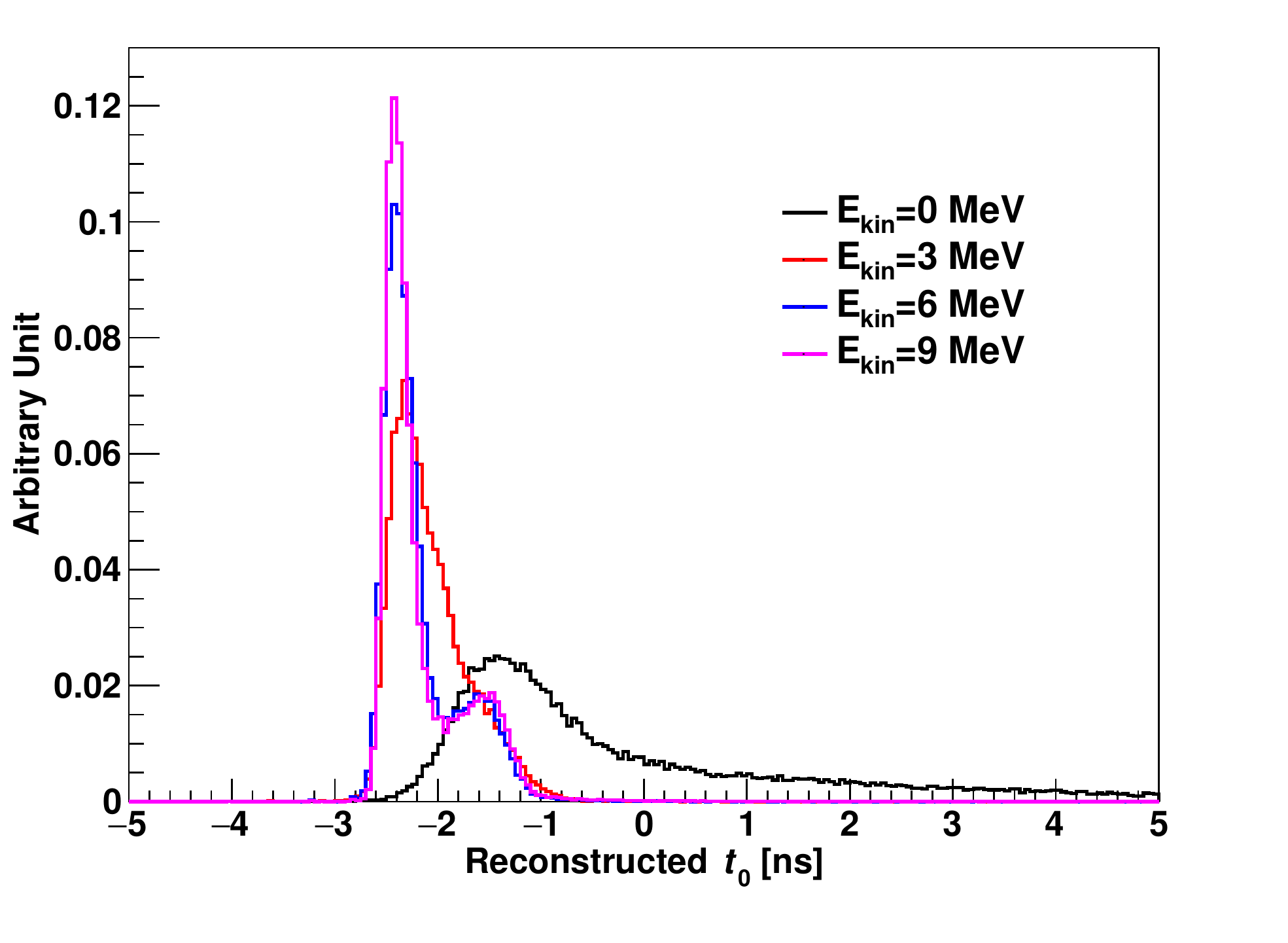}{\centering}		
		\caption{Reconstructed event time $t_{0}$ at different energies.}
		\label{fig:tlh_t0}			
	\end{figure}   

	The reconstructed event time $t_{0}$ is shown in Fig.~\ref{fig:tlh_t0}. The effect of $ t_ {0} $ is essentially a global shift of an event to match the scintillator response function PDF; in reality, $t_{0}$ is also affected by the trigger time and the time delay from the cable. The absolute value of $t_{0}$ can be neglected; only the relative difference of different events is important for the alignment of events. The small bump near $-$1.6~ns is correlated with the $R$ bias, and the long tail on the right side results from positronium formation.  The variation in the reconstructed $t_{0}$ is within a few nanoseconds.
	
\section{Total reflection region calculated by the charge likelihood algorithm}
\label{sec:clh}
	The time likelihood method described in Sec.~\ref{sec:tlh} introduces a bias in the $R$ direction when the reconstructing events are close to the acrylic sphere. As mentioned in Ref.~\cite{Smirnov_2003}, using a charge signal with the maximum likelihood method can provide better spatial resolution than the time likelihood algorithm when an event occurs near the detector boundary. In this section, we discuss the charge likelihood algorithm to reconstruct the event vertex in the total reflection region only, while the reconstruction result in the central region is omitted.

	The charge likelihood algorithm is based on the distribution of the number of photoelectrons in each PMT. With the mean expected number of photoelectrons $\mu(\vec{r_{0}},E)$ detected by each PMT at a given vertex and energy, the probability of observing $N_{pe}$ on a PMT follows a Poisson distribution. Furthermore,
	\begin{enumerate}
		\item[$*$] Probability for the $j$th PMT with no hits: $P_{nohit}^{j}(\vec{r_{0}},E) = e^{-\mu_{j}}$,
		\item[$*$] Probability for the $i$th PMT with $N^{i}_{pe}$ hits: $P_{hit}^{i}(\vec{r_{0}},E) = \frac{\mu_{i}^{N^{i}_{pe}}e^{-\mu_{i}}}{N^{i}_{pe}!}$.
	\end{enumerate}

		Therefore, the probability of observing a hit pattern for an event can be written as
		\begin{equation}
			p(\vec{r_{0}},E) =\prod_{j}{P_{nohit}^{j}(\vec{r_{0}},E)}\cdot{\prod_{i}{P_{hit}^{i}(\vec{r_{0}},E)}}.
			\label{equ:clhprob}
		\end{equation}		

	 The best-fit values of $\vec{r_{0}}$ and $E$ can be obtained by minimizing the negative log-likelihood function 
		\begin{equation}
			\mathcal{L} {(\vec{r_{0}},E)} = -\ln(p(\vec{r_{0}},E)).
		\end{equation}	
 
 	In principle, $\mu(\vec{r_{0}},E)$ can be expressed by the equation

	\begin{equation}
		\mu_{i}(\vec{r_{0}},E) = Y\cdot\frac{\Omega(\vec{r_{0}},r_{i})}{4\pi}\cdot\varepsilon_{i}\cdot{f(\theta_{i})}\cdot{e^{-\sum_{m}{\frac{d_{m}}{\zeta_{m}}}}}\cdot{E} + \delta_{i},		 
		 \label{equ:expnpe}
	\end{equation}	
	where $Y$ is the energy scale factor, $\Omega(\vec{r_{0}},r_{i})$ is the solid angle of the $i$th PMT, $\varepsilon_{i}$ is the detection efficiency of the $i$th PMT, $f(\theta_{i})$ is the angular response of the $i$th PMT, $\theta_{i}$ is defined in Fig.~\ref{fig:lightpath}, $\zeta_{m}$ is the attenuation length~\cite{attleng} in materials, and $\delta_{i}$ is the expected number of dark noise. This equation is based on the assumption that the scintillation light yield is linearly proportional to the energy. 

	\begin{figure}[!ht]
		\centering
		\includegraphics[width=0.5\textwidth]{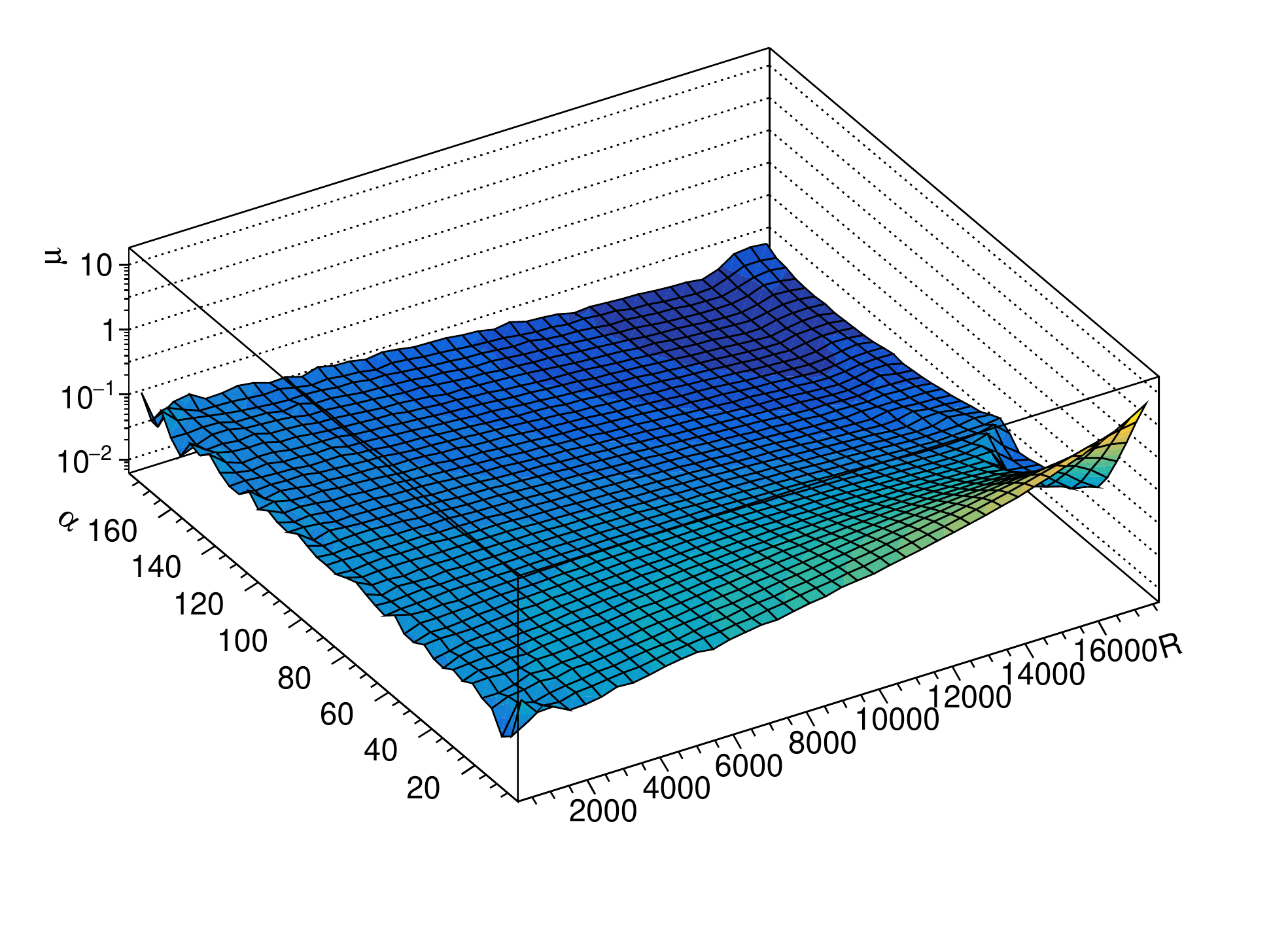}{\centering}		
		\caption{Mean expected number of photoelectron distribution as a function of radius $R$ and angle $\alpha$. This map is obtained by placing gamma sources at 29 specific positions along the Z-axis, which can be performed using a calibration procedure~\cite{Wu_2019}.}
		\label{fig:clh_charge_map}			
	\end{figure}   		
	
	However, Eq.~\ref{equ:expnpe} cannot describe properly the contribution of the indirect light, the effect of light shadows because of the geometric structure, and the effect of the total reflection. Another solution is to use the model-independent method described in Ref.~\cite{Wu_2019}: the mean expected number of photoelectrons can be obtained by placing gamma sources at 29 specific positions along the Z-axis, which can be performed using a calibration procedure~\cite{calibsys}. In this study, different from Ref.~\cite{Wu_2019}, we focused on the performance of the vertex reconstruction. The mean expected number of the photoelectron distributions as a function of radius $R$ and angle $\alpha$ is shown in Fig.~\ref{fig:clh_charge_map}, and the definition of angle $\alpha$ is shown in Fig.~\ref{fig:lightpath}.

	The mean expected number of photoelectrons $\mu$ obtained from Fig.~\ref{fig:clh_charge_map} was used to calculate the hit probability. Instead of reconstructing ($R,\theta,\phi$) at the same time, $\theta$ and $\phi$ were fixed at the reconstructed values provided by the time likelihood algorithm, and only the event radius $R$ was reconstructed using the charge likelihood algorithm. Therefore, the probability in Eq.~\ref{equ:clhprob} can be rewritten as
		\begin{equation}
			p(R,E) =\prod_{j}{P_{nohit}^{j}(R,E)}\cdot{\prod_{i}{P_{hit}^{i}(R,E)}}.
		\end{equation}		
	
\begin{figure}[!ht]
		\centering
		\includegraphics[width=0.5\textwidth]{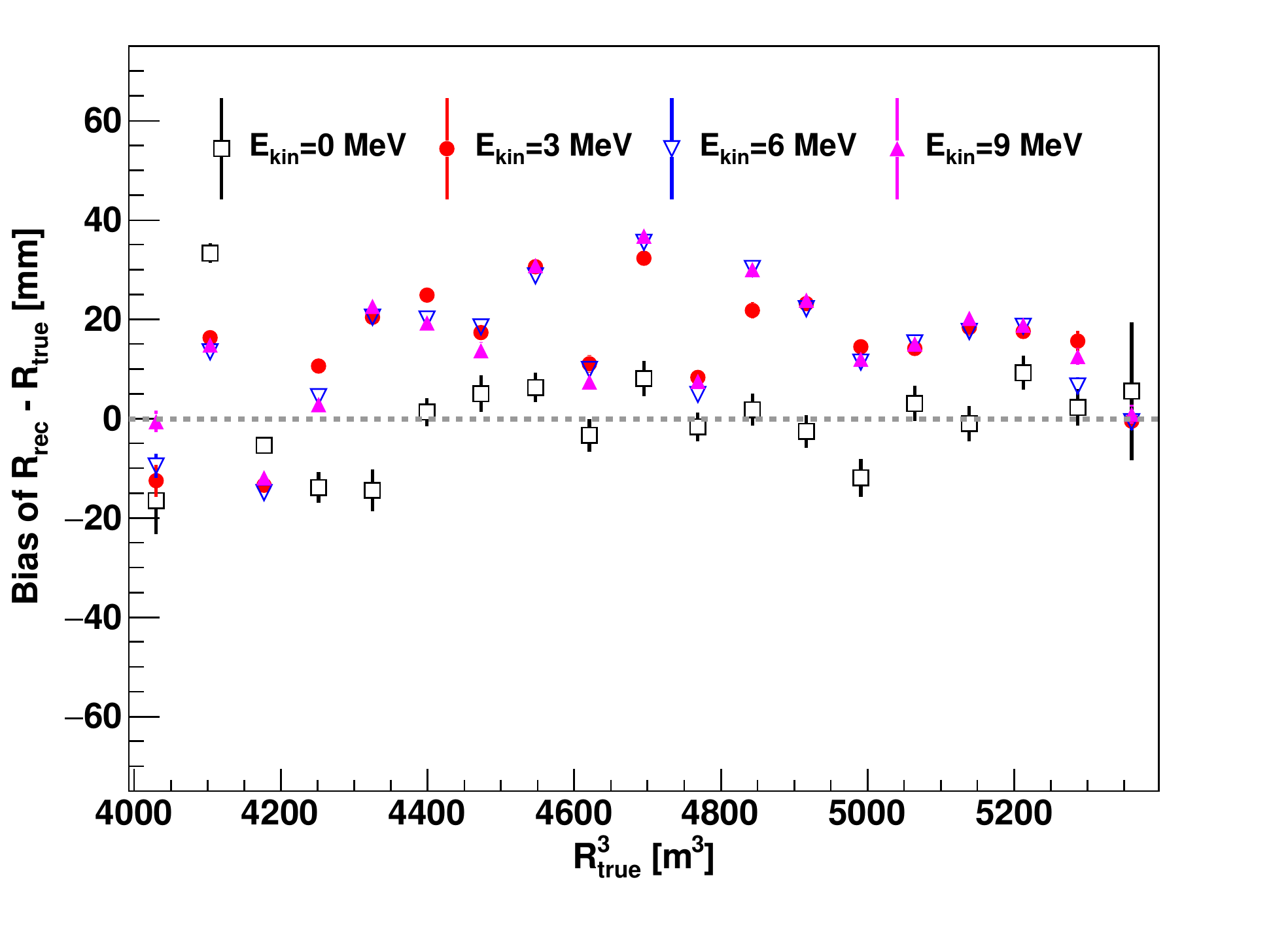}{\centering}		
		\caption{Bias of the reconstructed $R$ in the total reflection region at different energies calculated by the charge likelihood algorithm.}
		\label{fig:clh_bias}			
	\end{figure}

\begin{figure}[!ht]
		\centering
		\includegraphics[width=0.5\textwidth]{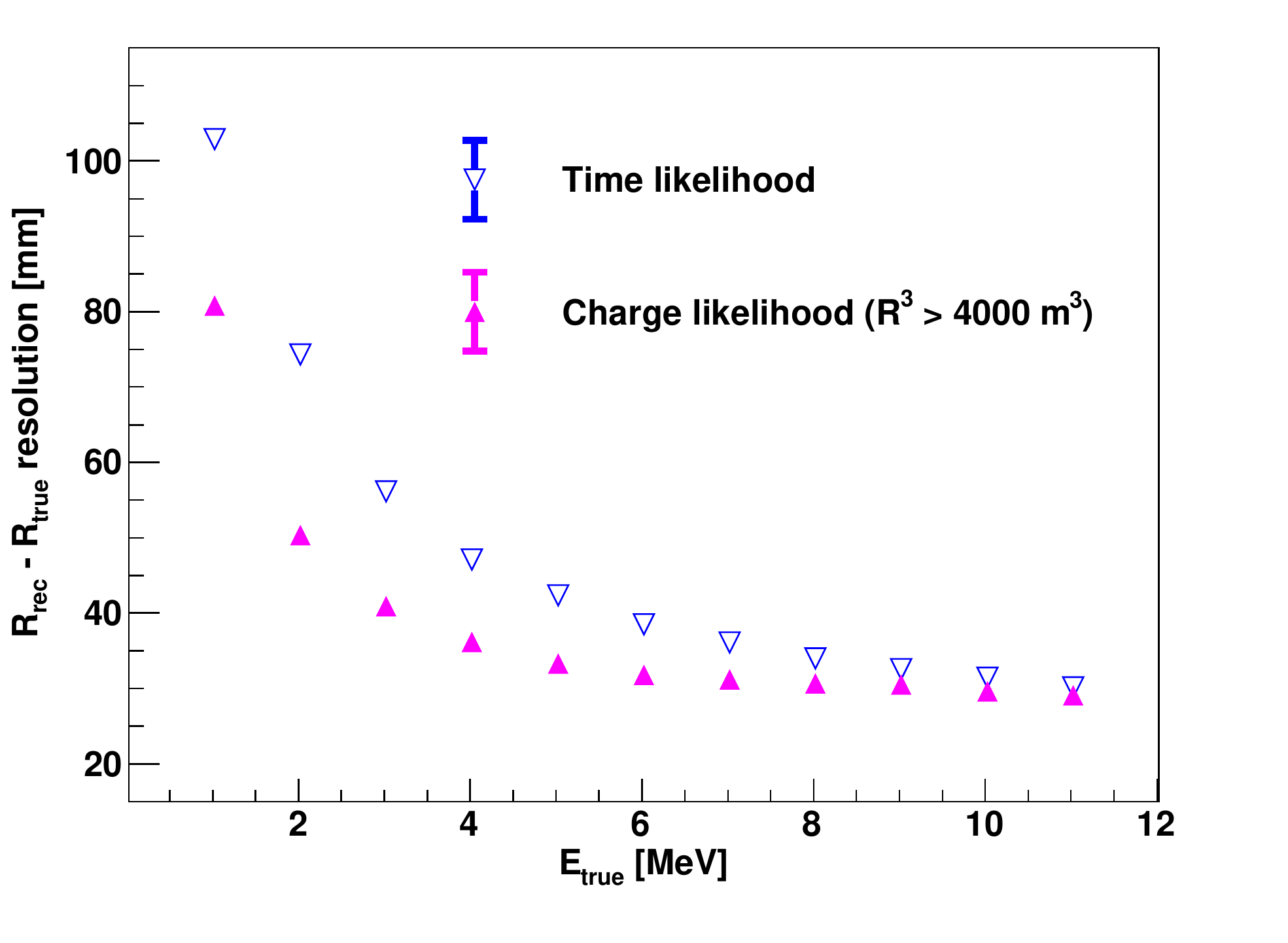}{\centering}		
		\caption{Resolution of reconstructed $R$ as a function of energy calculated by the time likelihood and charge likelihood algorithms ($R^{3}~>~4000~\rm{m}^{3}$)}.
		\label{fig:clh_res}			
	\end{figure}

	The reconstruction performance, focusing on the total reflection region, is shown in Figs.~\ref{fig:clh_bias} and~\ref{fig:clh_res}. In the total reflection region, the mean value of the reconstructed $R$ was consistent with the true $R$, and the resolutions in the $R$ direction were 81~mm at 1.022~MeV and 30~mm at 10.022~MeV. 

	As the charge distribution provides good radial discrimination ability, this algorithm can provide better resolution and a significantly smaller bias compared with those of the time likelihood algorithm in the total reflection region.
	
\section{Performance summary}
\label{sec:summary}

\begin{figure}[!ht]
		\centering
		\includegraphics[width=0.5\textwidth]{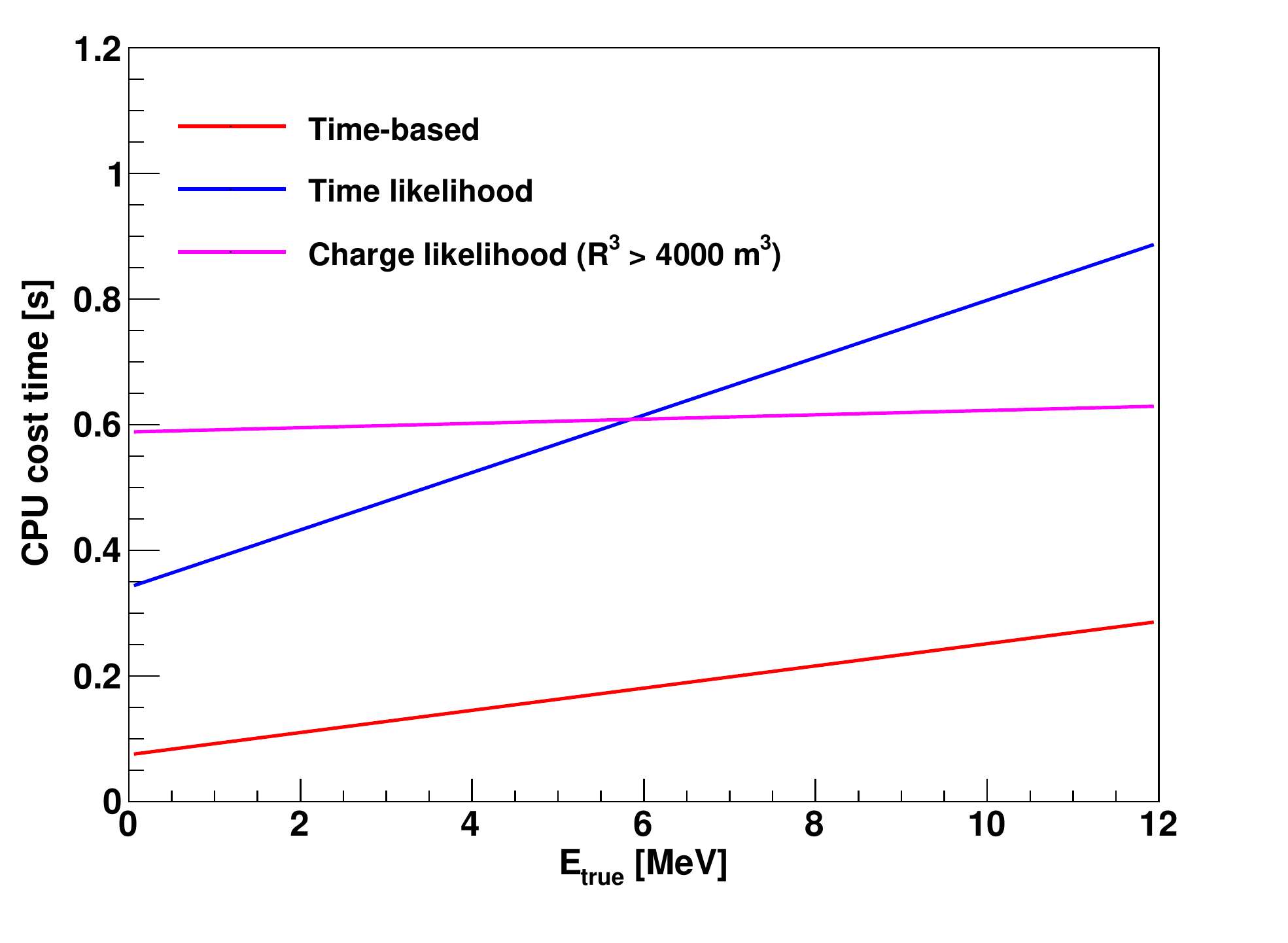}{\centering}		
		\caption{Execution time for the reconstruction for different algorithms.}
		\label{fig:cputime}			
	\end{figure}

	The execution time of the reconstruction for each event was tested on a computing cluster with Intel Xeon Gold 6238R CPUs (2.2~GHz), as shown in Fig.~\ref{fig:cputime}. The execution time of the charge-based algorithm was in the order of $O(10^{-4})$~s per event, which cannot be presented in the figure. The execution time of the time-based and the time likelihood algorithm was proportional to the event energy and could be reduced by using only the Hamamatsu PMTs for the reconstruction. The execution time of the charge likelihood algorithm was independent of the event energy.
	
\begin{figure}[!ht]
		\centering
		\includegraphics[width=0.5\textwidth]{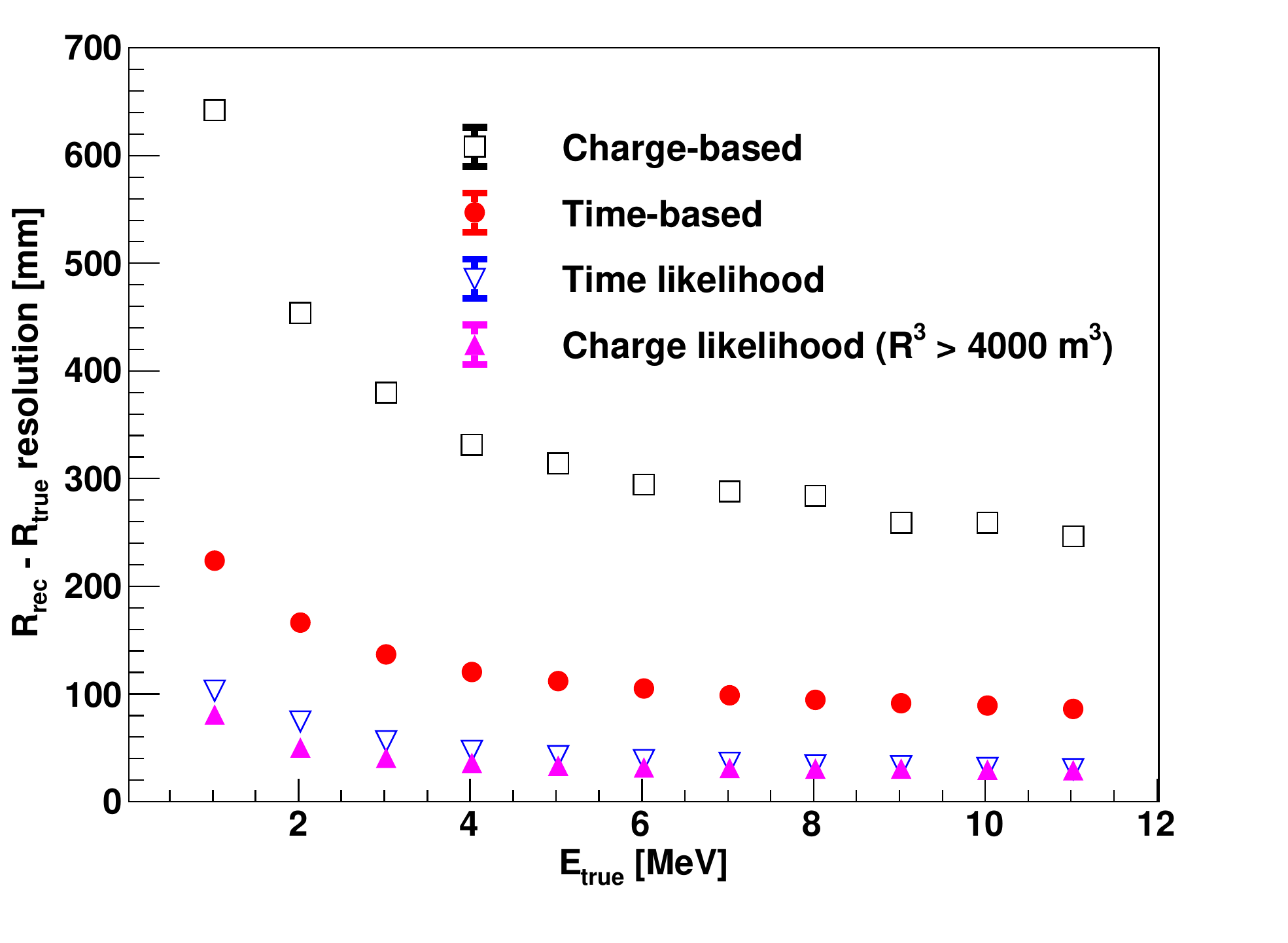}{\centering}		
		\caption{Resolution of the reconstructed $R$ as a function of energy for different algorithms.}
		\label{fig:algrescomp}			
	\end{figure}

	The resolutions of the four algorithms in the $R$ direction are shown in Fig.~\ref{fig:algrescomp}. Owing to the large bias of the charge-based algorithm, a correction to remove the position-dependent bias was applied before the analysis of the resolution.
	
	The charged-based algorithm is suitable for online reconstruction tasks that require high speed but do not require high resolution. The time-based algorithm does not rely on MC; it can be used as a data-driven reconstruction method. The time likelihood and charge likelihood algorithms are relatively accurate and each has its own advantages in a specific detector region.

\section{Discussion}	
\label{sec:discussion}
	The vertex resolutions of KamLAND and Borexino are approximately 12~cm and 10~cm at 1~MeV, respectively, and for JUNO it is approximately 10.5~cm. The diameter of JUNO (35.4~m) is several times larger than that of KamLAND (13~m) and Borexino (8.5~m). Despite its larger size, JUNO is still able to achieve a similar vertex resolution based only on the PMT time information. In this study, various effects on the vertex reconstruction for JUNO were comprehensively analyzed. As expected, the TTS of the PMT is the dominant factor. The vertex reconstruction capability of JUNO is mainly based on the Hamamatsu PMTs. Although the number of NNVT PMTs is more than twice, their time information is not useful in the vertex reconstruction because of their significantly inferior TTS. After considering the light yield of the LS and the PMT coverage, the number of photons detected by the Hamamatsu PMTs is in the same range as those of the three detectors mentioned above. This provides an explanation of their similar vertex resolutions based only on the PMT time information. To fully exploit the large PMT coverage and large number of PMTs in JUNO, the charge information of PMTs also need to be utilized in addition to the time information to constrain the event vertex, especially near the detector boundary. At the same time, the effect of the dark noise of the PMT can be mitigated with appropriate treatment. A more accurate initial value also improves the performance of the vertex reconstruction. In addition to the event vertex, the event time can also be reconstructed simultaneously, which is a useful variable for downstream analyses.

	In LS detectors, in addition to the scintillation photons there are also Cherenkov photons, whose effects need to be studied in the future. The $R$ bias near the detector edge in the current results also indicates that a more accurate PDF of the scintillator response function is needed to include its dependence on the position as well as the particle type. All vertex reconstruction methods based on PMT time information use the time of the first arrival photons only; in principle, later photons might be useful as well. Therefore, novel methods also need to be explored. Our preliminary studies on algorithms based on machine learning~\cite{qianz_2021} showed comparable vertex reconstruction performance, and they need to be further investigated 

\section{Conclusion}	
\label{sec:conclusion}
	In this study, four algorithms for the reconstruction of the event vertex and event time were investigated in detail and verified using MC samples generated by the offline software of JUNO. Considering the TTS and dark noise effects from the PMTs, a vertex resolution of 10~cm or higher can be achieved in the energy range of reactor neutrinos. The TTS has a major effect on the vertex resolution, whereas the effect of dark noise is limited. Near the detector boundary, charge information can constrain the event vertex better compared with time information. The algorithms discussed in this paper are applicable to current and future experiments using similar detection techniques.

\end{document}